\documentclass[lettersize,journal]{IEEEtran}

\usepackage[T1]{fontenc}
\usepackage[utf8]{inputenc} 
\usepackage{graphicx}
\usepackage[]{hyperref}
\usepackage[version=3]{mhchem}
\usepackage{amssymb}
\usepackage{longtable}
\usepackage{multirow}
\usepackage{amsmath}
\usepackage{eurosym}	 
\usepackage{mathtools}
\usepackage{booktabs}

\usepackage{upgreek}

\usepackage{epstopdf}
\usepackage{caption}
\usepackage{color}
\usepackage{xcolor}
\definecolor{myred}{rgb}{0.7,0.15,0.15}
\definecolor{mygreen}{rgb}{0.13,0.55,0.13}
\definecolor{myblue}{rgb}{0.25,0.41,0.88}

\usepackage{graphicx}
\usepackage[labelformat=simple]{subcaption} 

\usepackage{placeins}
\usepackage[font={small}]{caption}

\usepackage{mathrsfs}

\usepackage{enumitem}
\usepackage{url}
\usepackage{times}

\usepackage{pgfplots}
\usepgfplotslibrary{groupplots}
\usepackage{tikz}
\usetikzlibrary{patterns}
\usetikzlibrary{external}
\usetikzlibrary{spy}

\usepackage{pgfkeys}
\usetikzlibrary{intersections}
\usepackage{listings}

\newlength\Colsep
\DeclareMathOperator{\atantwo}{atan2}

\newcommand{\tdp}{2D-$\xi$}
\newcommand{\tdpk}{2D-$\xi$}

\newcommand{\ezh}{\hat{\vec e}_{\xi_3}}

\newcommand{\ey}{\hat{\vec e}_{y}}
\newcommand{\ez}{\hat{\vec e}_{z}}

\renewcommand{\vec}[1]{\boldsymbol{#1}}
\newcommand{\mat}[1]{\mathbf{#1}}
\newcommand{\der}[2]{\frac{\partial #1}{\partial #2}}

\newcommand{\derd}[2]{\frac{d #1}{d #2}}
\newcommand{\paren}[1]{\left(#1\right)}

\newcommand{\volInt}[3]{\paren{#1 \, , #2}_{#3}}

\newcommand{\grad}{\textrm{\textbf{grad}}\, }

\renewcommand{\div}{\textrm{div}\, }

\newcommand{\curl}{\textrm{\textbf{curl}}\, }
\newcommand{\curlOnly}{\textrm{\textbf{curl}}}

\renewcommand{\b}{\vec b}

\newcommand{\n}{\vec n}
\newcommand{\h}{\vec h}

\newcommand{\e}{\vec e}

\renewcommand{\j}{\vec j}

\newcommand{\dt}{\partial_t}
\newcommand{\ec}{e_{\text{c}}}
\newcommand{\jc}{j_{\text{c}}}

\renewcommand{\O}{\Omega}

\newcommand{\Oci}{\Omega_{\text{c}_i}}

\newcommand{\Oc}{\Omega_{\text{c}}}
\newcommand{\Occ}{\Omega_{\text{c}}^{\text{C}}}

\newcommand{\Gout}{\Gamma_{\text{out}}}

\newcommand{\mutilde}{\tilde{\vec \mu}}
\newcommand{\rhotilde}{\tilde{\vec \rho}}
\newcommand{\Othreed}{\O_{\text{3D}}}
\newcommand{\Octhreed}{\O_{\text{c,3D}}}

\newcommand{\hpe}{\vec h_{\perp}}
\newcommand{\hpa}{\vec h_{\parallel}}
\newcommand{\hpek}[1]{\vec h_{\perp , #1}}

\newcommand{\hpak}[1]{\vec h_{\parallel , #1}}
\newcommand{\phipek}[1]{h_{\perp , #1}}
\newcommand{\phipak}[1]{\phi_{\parallel , #1}}

\newcommand{\hsppa}{\mathcal{H}_{\parallel}}
\newcommand{\hsppe}{\mathcal{H}_{\perp}}
\newcommand{\hspzpa}{\mathcal{H}_{\parallel, 0}}
\newcommand{\hspzpe}{\mathcal{H}_{\perp, 0}}

\newcommand{\hpf}{$h$-$\phi$-formulation\ }

\newcommand{\hfOnly}{$h$-formulation}
\newcommand{\hpfOnly}{$h$-$\phi$-formulation}

\newcommand{\hsp}{\mathcal{H}}

\newcommand{\hspz}{\mathcal{H}_{0}}

\newcommand{\oneforms}{$1$-forms}

\newcommand{\nodes}{\mathcal{N}}
\newcommand{\edges}{\mathcal{E}}

\newcommand{\transpose}{^{\text T}}
\newcommand{\inv}[1]{\mat #1^{-1}}

\usepackage{eso-pic}

\newcommand{\sinc}{\mathrm{s}}
\newcommand{\cosc}{\mathrm{c}}

\usepackage{color}
\usepackage{xcolor}
\definecolor{myred}{rgb}{0.7,0.15,0.15}
\definecolor{mymaincolor}{rgb}{0.24, 0.36, 0.64}
\definecolor{mysecondcolor}{rgb}{0.21, 0.64, 0.87}
\definecolor{myblue}{rgb}{.2,0.45,0.5} 
\definecolor{myorange}{rgb}{0.78,0.6,0.3}
\definecolor{mygreen}{rgb}{.2,0.38,0.16}
\definecolor{myalert}{rgb}{0.97,0.09,0.21}

\definecolor{vir_0}{rgb}{0.993248, 0.906157, 0.143936}
\definecolor{vir_1}{rgb}{0.565498, 0.84243 , 0.262877}
\definecolor{vir_2}{rgb}{0.20803 , 0.718701, 0.472873}
\definecolor{vir_3}{rgb}{0.128729, 0.563265, 0.551229}
\definecolor{vir_4}{rgb}{0.190631, 0.407061, 0.556089}
\definecolor{vir_5}{rgb}{0.267968, 0.223549, 0.512008}
\definecolor{vir_6}{rgb}{0.267004, 0.004874, 0.329415}

\definecolor{mag_0}{rgb}{0.001, 0 , 0.014}
\definecolor{mag_1}{rgb}{0.135, 0.068, 0.315}
\definecolor{mag_2}{rgb}{0.372, 0.093, 0.499}
\definecolor{mag_3}{rgb}{0.595, 0.176, 0.501}
\definecolor{mag_4}{rgb}{0.829, 0.262, 0.431}
\definecolor{mag_5}{rgb}{0.973, 0.462, 0.362}
\definecolor{mag_6}{rgb}{0.997, 0.734, 0.505}
\definecolor{mag_7}{rgb}{0.987, 0.991, 0.75 }

\definecolor{myformulation}{rgb}{0.33, 0.29, 0.31}
\definecolor{myformulation_back}{rgb}{1, 0.97, 0.91}

\definecolor{hf}{rgb}{0.93, 0.57, 0.13} 
\definecolor{hf_2}{rgb}{1.0, 0.89, 0.77} 
\definecolor{hf_3}{rgb}{1.0, 0.22, 0.0} 
\definecolor{hf_4}{rgb}{1.0, 0.4, 0.1} 

\definecolor{burlywood}{rgb}{0.87, 0.72, 0.53}
\definecolor{burntorange}{rgb}{0.8, 0.33, 0.0}
\definecolor{burntsienna}{rgb}{0.91, 0.45, 0.32}

\definecolor{af}{rgb}{0.4, 0.53, 0.34}
\definecolor{af_2}{rgb}{0.74, 0.77, 0.47}
\definecolor{af_3}{rgb}{0.12, 0.3, 0.17}
\definecolor{af_4}{rgb}{0.03, 0.34, 0.25}

\definecolor{haf}{rgb}{0.6, 0.51, 0.48}
\definecolor{haf_2}{rgb}{1, 0.97, 0.91}

\definecolor{taf}{rgb}{0, 0.55, 0.5}
\definecolor{ajf}{rgb}{0.29, 0.59, 0.82}

\definecolor{hbf}{rgb}{0.87, 0.36, 0.51}

\definecolor{prussianblue}{rgb}{0.0, 0.19, 0.33}
	\definecolor{regalia}{rgb}{0.32, 0.18, 0.5}
\begin{document}

\title{Helicoidal Transformation Method for Finite Element Models of Twisted Superconductors}

\author{Julien~Dular, François~Henrotte, André~Nicolet,
        Mariusz~Wozniak, Benoît~Vanderheyden,~and~Christophe~Geuzaine
\thanks{J. Dular and M. Wozniak are with CERN, Geneva, Switzerland. A. Nicolet is with the Aix-Marseille University, France. F. Henrotte, B. Vanderheyden, and C. Geuzaine are with the University of Liège, Belgium.}%
}

\maketitle

\begin{abstract}
This paper deals with the modelling of superconducting and resistive wires with a helicoidal symmetry, subjected to an external field and a transport current. Helicoidal structures are three-dimensional, and therefore yield computationally intensive simulations in a Cartesian coordinate system. We show in this paper that by working instead with a helicoidal system of coordinates, the problem to solve can be made two-dimensional, drastically reducing the computational cost. We first introduce the state-of-the-art approach and apply it on the \hpf with helicoidally symmetric boundary conditions (e.g., axial external magnetic field, with or without transport current), with an emphasis on the function space discretization. Then, we extend the approach to general boundary conditions (e.g., transverse external magnetic field) and we present numerical results with linear materials. In particular, we discuss the frequency-dependent losses in composite wires made of superconducting filaments embedded in a resistive matrix. Finally, we provide outlook to the application of the generalized model with nonlinear materials.
\end{abstract}
%



\AddToShipoutPicture*{
    \footnotesize\sffamily\raisebox{0.8cm}{\hspace{1.5cm}\fbox{
        \parbox{\textwidth}{
            This work has been submitted to IEEE for possible publication. Copyright may be transferred without notice, after which this version may no longer be accessible.
            }
        }
    }
}

\section{Introduction}
\addcontentsline{toc}{section}{Introduction}

\IEEEPARstart{L}{ow-temperature} superconducting composite wires usually consist of a large number of superconducting filaments embedded in a conducting matrix. This matrix helps in redistributing current between filaments, but has the side effect of coupling the filaments in the presence of an external transverse time-varying magnetic field. This coupling can however be reduced by twisting the composite wire~\cite{carr1974ac,wilson1983superconducting}. The resulting geometry is not invariant along the wire axis and leads to a computationally intensive three-dimensional (3D) modelling~\cite{zhao20173d,escamez20163,escamez2017experimental,riva2023h}.

Approximate models exploiting the multifilamentary structure of this kind of wires have been investigated to reduce the computational cost, such as in \cite{satiramatekul2010numerical} and \cite{satiramatekul2005contribution}, where coupling currents in the conducting matrix are accounted for in a 2D finite element model by introducing equivalent 
resistances between the filaments. Alternatively, a Frenet frame is used in~\cite{kameni2019reduced} to simplify the definition of the 3D geometry, and AC losses are approximated by considering a fraction of the pitch length of the wire in a 3D model, or a cross section of the wire in a 2D model. Homogenization techniques involving anisotropic materials have also been considered~\cite{zhao20173d}. Finally, parallelization methods are considered to reduce the computational time~\cite{riva2023h}.

Whenever possible, it is always recommended to exploit existing symmetries. In particular, the dimension of a problem presenting a helicoidal symmetry, i.e., a combination of translational and rotational symmetries with the same axis, can be reduced from 3D to 2D without loss of accuracy if the calculations are performed in a \textit{helicoidal coordinate system}.
Methods based on this coordinate transformation have first been introduced in optical waveguide simulations~\cite{nicolet2004modelling,nicolet2006finite, nicolet2007leaky}, and since then applied to electrostatic problems \cite{nicolet2006asymptotic, hazim20212d}, linear magnetodynamic problems~\cite{piwonski20232d, marjamaki2023utilizing, piwonski2023_aform}, and nonlinear magnetodynamic problems with superconducting filaments or tapes~\cite{stenvall2013computation, lahtinen2013toward, stenvall2013manifolds, dular2023standard}. 

An exact helicoidal symmetry is rarely encountered in practical applications, but different kinds of deformed geometries, curved wires, or conductor organized, e.g., into layers with distinct twist pitch lengths, may exhibit an approximate or partial helicoidal symmetry. Working with helicoidal coordinate systems can still be very useful in such cases, especially in the context of a multi-scale or a sub-problem approach, to compute homogenized parameters that account for the twisting of the filaments (e.g., Rutherford multistrand cables). Furthermore, the 2D helicoidal approach is more accurate than an equivalent 3D approach, as the latter is usually limited in accuracy by non-conformities at element interfaces in unstructured 3D meshes. Extensions and improvements of the helicoidal method are therefore currently being investigated, e.g., in~\cite{marjamaki2023utilizing}, to quantify helicoidal effects in the context of Litz wires.

This paper focuses on the helicoidal transformation method. We start the analysis in Section~\ref{sec_helicoidal_change} by applying the change of coordinates to the \hpfOnly~\cite{bossavit1998computational}, which is an efficient formulation for systems with superconductors~\cite{dular2019finite}, and we then state the mathematical conditions for reducing the problem dimension from 3D to 2D. We will refer to the equations resulting from this analysis as the \tdp\ model, in order to emphasize the fact that it is solved in helicoidal coordinates. As will be shown, a feature of the \tdp\ model, compared to a conventional 2D model in Cartesian coordinates, is that it solves for fields with three independent components, instead of two.

Depending on the symmetry of the boundary conditions (BC), the study is decomposed in two cases. If the magnetic field excitation is axial and uniform, the BC then also verify the helicoidal symmetry, irrespective of whether there is a transport current or not. The dimension of this problem (geometry plus BC), being helicoidally symmetric, can be reduced from 3D to 2D by simply applying the coordinate transformation method. This approach is not new in the context of superconducting wires~\cite{stenvall2013computation, lahtinen2013toward, stenvall2013manifolds}, but it has not yet been presented with the efficient \hpfOnly. In Section~\ref{sec_helicoidal_symmetric2D}, the implementation details of this formulation are reviewed with an emphasis on the discretization of a curl-free magnetic field in non-conducting domains.
In Section~\ref{sec_filaments_verif3Dmodel}, the implementation is verified by comparison with a 3D model in Cartesian coordinates.

If the magnetic field excitation is transverse, then the BC are no longer helicoidally symmetric. A generalization of the method is proposed in Section~\ref{sec_helicoidal_non_symmetric} for this case. It still results in a 2D model in some situations. To the best of our knowledge, this generalization is a novelty compared to state-of-the-art methods. Attention is again paid to the curl-free property of the magnetic field in non-conducting domains. The generalized model is applied to linear materials in Section~\ref{sec_filaments_verifquasi3Dmodel}, and it is shown that it reproduces the predictions of analytical models for the coupling currents~\cite{wilson1983superconducting}. We finally provide a brief prospect about the application of the generalized method in the presence of nonlinear materials.

All presented models are implemented in and solved by GetDP~\cite{getdp}. Geometry and mesh generation are performed by Gmsh~\cite{gmsh}. All codes are open-source and available in the Life-HTS toolkit\footnote{Available: \url{www.life-hts.uliege.be}.}.

\section{Helicoidal Change of Coordinates}\label{sec_helicoidal_change}

Let $(x,y,z)$ be a Cartesian coordinate system. 
The \textit{helicoidal change of coordinates} $\vec x \to \vec \xi$
and its inverse $\vec \xi \to \vec x$ read~\cite{nicolet2006finite}
\begin{align}\label{eq_helicoidalCV}
&\left\{\begin{aligned}
\xi_1 &= x \cos(\alpha z) + y \sin(\alpha z),\\
\xi_2 &= -x \sin(\alpha z) + y \cos(\alpha z),\\
\xi_3 &= z,
\end{aligned}\right.
\end{align}
and 
\begin{align}\label{eq_helicoidalCV_inv}
&\left\{\begin{aligned}
x &= \xi_1 \cos(\alpha \xi_3) - \xi_2 \sin(\alpha \xi_3),\\
y &= \xi_1 \sin(\alpha \xi_3) + \xi_2 \cos(\alpha \xi_3),\\
z &= \xi_3,
\end{aligned}\right.
\end{align}
respectively, 
with $(\xi_1, \xi_2, \xi_3)$ the helicoidal coordinate system.
The twisting parameter $\alpha \in \mathbb{R}$ is the unique parameter of the coordinate transformation, and the pitch length is $p = 2\pi /\alpha$. 

\begin{figure}[h!]
         \centering
        \begin{subfigure}[b]{0.49\linewidth}
            \centering
            \tikzsetnextfilename{helicoidalTransfo_simple_a}
	\begin{tikzpicture}[trim axis left, trim axis right][font=\small]
 	\begin{axis}[
    width=1.1\linewidth,
    height=5cm,
    grid = both,
    grid style = dotted,
    view={65}{30},
    zmin=0, 
    zmax=1,
    xmin=-1, 
    xmax=1,
    ymin=-1, 
    ymax=1,
	xlabel={$x$},
    ylabel={$y$},
    zlabel={$z$},
    xlabel style={xshift=0.3em, yshift=1.5em},
    ylabel style={yshift=1.2em},
    zlabel style={xshift=-0.5em,yshift=-0.5em},
    ]
\addplot3+[domain=0:2*pi,samples=60,samples y=0, no marks, very thick, gray]
({0.9*cos(deg(x))},
{0.9*sin(deg(x))},
{x/(2*pi)});
\addplot3+[domain=0:2*pi,samples=60,samples y=0, no marks, very thick, black]
({0.5*cos(deg(x))},
{0.5*sin(deg(x))},
{x/(2*pi)});
\end{axis}
	\end{tikzpicture}%
	\vspace{-0.2cm}
            \caption{Cartesian coordinates.}
            \label{helicoidalTransfo_simple_a}
        \end{subfigure}
        \hfill
        \begin{subfigure}[b]{0.49\linewidth}  
            \centering 
            \tikzsetnextfilename{helicoidalTransfo_simple_b}
	\begin{tikzpicture}[trim axis left, trim axis right][font=\small]
 	\begin{axis}[
    width=1.1\linewidth,
    height=5cm,
    grid = both,
    grid style = dotted,
    view={65}{30},
    zmin=0, 
    zmax=1,
    xmin=-1, 
    xmax=1,
    ymin=-1, 
    ymax=1,
	xlabel={$\xi_1$},
    ylabel={$\xi_2$},
    zlabel={$\xi_3$},
    xlabel style={xshift=0.3em, yshift=1.5em},
    ylabel style={yshift=1.2em},
    zlabel style={xshift=-0.5em,yshift=-0.5em},
    ]
\addplot3+[domain=0:2*pi,samples=60,samples y=0, no marks, very thick, gray]
({0.9},
{0},
{x/(2*pi)});
\addplot3+[domain=0:2*pi,samples=60,samples y=0, no marks, very thick, black]
({0.5},
{0},
{x/(2*pi)});
\end{axis}
	\end{tikzpicture}%
	\vspace{-0.2cm}
            \caption{Helicoidal coordinates.}
            \label{helicoidalTransfo_simple_b}
        \end{subfigure}
        \caption{Transformation of two helicoidal curves with the change of coordinates Eq.~\eqref{eq_helicoidalCV} with $\alpha = 2\pi$.}
        \label{helicoidalTransfo_simple}
\end{figure}
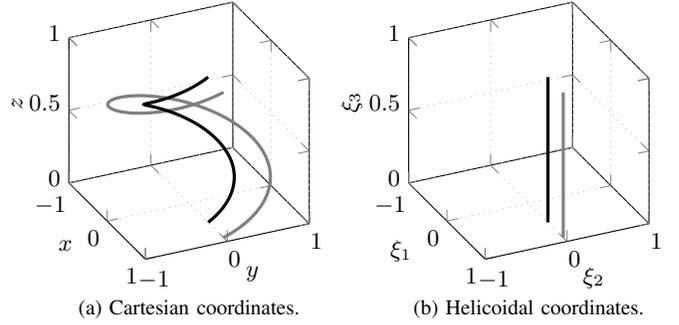

With this transformation, helices of pitch length $p$ around the $z$-axis in the Cartesian coordinate system are mapped into straight lines parallel to the $\xi_3$-axis in the helicoidal coordinate system. This is illustrated in Fig.~\ref{helicoidalTransfo_simple} with $p=1$. A geometry is said to be helicoidally symmetric, or to have a helicoidal symmetry, if there exists a value $\alpha$ for which its description in helicoidal coordinates is $\xi_3$-invariant, i.e., independent of $\xi_3$. 

The Jacobian matrix $\mat J$ of the coordinate transformation Eq.~\eqref{eq_helicoidalCV_inv} reads
\begin{align}\label{eq_jacobian}
\mat J = \der{x_i}{\xi_j} = \begin{pmatrix}
\cosc & -\sinc & -\alpha \xi_1 \sinc- \alpha \xi_2\cosc\\
\sinc & \cosc & \alpha \xi_1 \cosc - \alpha \xi_2\sinc\\
0 & 0 & 1
\end{pmatrix},
\end{align}
with $\sinc = \sin(\alpha \xi_3)$ and $\cosc = \cos(\alpha \xi_3)$. We have $\det \mat J = 1$. The inverse transposed Jacobian matrix $\mat J^{-\text{T}}$, written in terms of the $\vec \xi$-coordinates, then reads
\begin{align}\label{eq_jacobian_invTrans}
\mat J^{-\text{T}} = \der{\xi_j}{x_i} = 
\begin{pmatrix}
\cosc & -\sinc & 0\\
\sinc & \cosc & 0\\
\alpha \xi_2 & - \alpha \xi_1 & 1
\end{pmatrix}.
\end{align}

\subsection{Helicoidal transformation of fields}

The Jacobian matrix describes the mapping of vector components with the transformation. Components of one-forms, such as the magnetic field $\h$, follow the transformation~\cite{nicolet2004modelling, ould2007transformations}
\begin{align}\label{eq_oneformTransfo}
\h_{\vec x} = \mat J^{-\text{T}}\ \h_{\vec \xi},
\end{align}
where $\h_{\vec x}$ and $\h_{\vec \xi}$ denote the components of the field $\h$ in the Cartesian and helicoidal coordinate systems, respectively. 

Components of two-forms, such as the current density $\j$ ($=\curl \h$), follow the transformation~\cite{nicolet2004modelling, ould2007transformations}
\begin{align}\label{eq_twoformTransfo}
\j_{\vec x} = \frac{\mat J}{\det \mat J}\ \j_{\vec \xi},
\end{align}
where $\j_{\vec x}$ and $\j_{\vec \xi}$ denote the components of the field $\j$ in the Cartesian and helicoidal coordinate systems, respectively. 

\subsection{Problem definition and \hpf}

The eddy current problem is governed by the following magnetodynamic (or magneto-quasistatic) equations and constitutive laws~\cite{jackson1999classical}:
\begin{equation}\label{MQSequations}
\left\{\begin{aligned}
\div\b &= 0,\\
\curl\h &= \j,\\
\curl\e &= -\dt \b,
\end{aligned}\right. \quad \text{and} \quad  \left\{\begin{aligned}
\b &= \mu\, \h,\\
\e &= \rho\, \j,
\end{aligned}\right.
\end{equation}
with $\b$, $\h$, $\j$ $\e$, $\mu$, and $\rho$, the magnetic flux density (T), the magnetic field (A/m), the current density (A/m$^2$), the electric field (V/m), the permeability (H/m), and the resistivity ($\O$m), respectively. In non-conducting materials, $\rho \to \infty$ and $\j = \vec 0$, and Ampère's law reads
\begin{align}\label{eq_curlFreeCondition}
\curl \h = \vec 0.
\end{align}
In this paper, special attention is paid to satisfy this condition.

Type-II irreversible superconductors are characterized by a nonlinear electric response. Assuming isotropy for low-temperature superconductors, their resistivity is given by the power law~\cite{rhyner1993magnetic},
\begin{equation}\label{eqn_contitutiveje}
\rho_\text{SC} = \frac{\ec}{\jc}\paren{ \frac{\|\j\|}{\jc}}^{n-1},
\end{equation}
where $\ec = 10^{-4}$ V/m is an electric field threshold defining the critical current density $\jc$ (A/m$^2$), and $n$ (-) describes the sharpness of the transition to flux flow. The norm of $\j$ is denoted by $\|\j\|$, with $\|\j\|^2 = j_x^2+j_y^2+j_z^2$ in the Cartesian coordinates.
Finally, all materials are assumed to be non-magnetic, so that one has $\mu = \mu_0 = 4\pi\times 10^{-7}$ H/m in all domains.

\begin{figure}[h!]
\begin{center}
\includegraphics[width=0.6\linewidth]{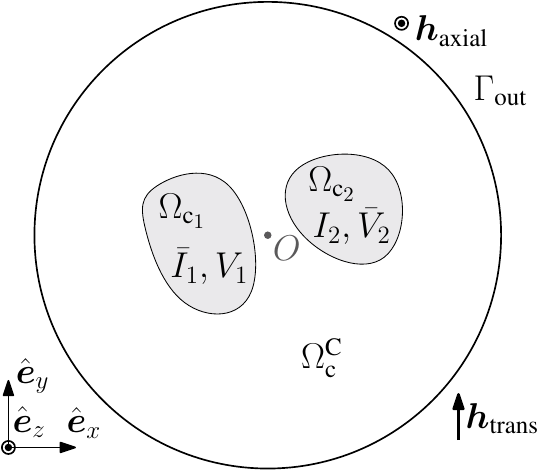}
\caption{2D cross section of the problem. Boundary conditions (BC) are axial magnetic field $\h_\text{axial}$ and transverse magnetic field $\h_\text{trans}$ imposed on $\Gout$. Global conditions (GC) are applied transport current $\bar I_1$ on $\O_{\text{c}_1}$, and applied voltage $\bar V_2$ on $\O_{\text{c}_2}$. In this example, $C = \{1,2\}$, $C_I = \{1\}$, and $C_V = \{2\}$. The 3D geometry is the rotated extrusion of the represented 2D cross section.}
\label{2D_generalTwistedProblem}
\end{center}
\end{figure}

The magnetodynamic problem defined above is solved in a computational domain $\O$. Let $\O$ be a  helicoidally symmetric domain. It consists of a conducting domain $\Oc$ made of $N$ connected subdomains, $\Oc=\cup_{i\in C} \Oci$, with $C = \{1,\dots,N\}$, surrounded by a non-conducting domain $\Occ$. The external boundary of $\O$ is noted $\Gout$. Via boundary conditions (BC), the system can be subjected to a given axial magnetic field $\h_{\text{axial}}$ and/or a given transverse magnetic field $\h_\text{trans}$. 
A transport current $\bar I_i$ is imposed to the subdomains $\Oci$ for $i\in C_I \subset C$, and a voltage $\bar V_i$ is imposed on the subdomains $\Oci$ for $i\in C\setminus C_I = C_V$. We shall call {\em global conditions} (GC) these electric conditions imposed to the conductors of the system. Fig.~\ref{2D_generalTwistedProblem} represents a typical cross section of the problem at hand. 

We solve the problem defined above with the finite element method. Among the existing finite element formulations, we choose the \hpfOnly~\cite{bossavit1998computational}. It involves the power law written in terms of the resistivity, which has been shown to lead to robust and efficient numerical resolutions for problems involving superconductors characterized by the power law~\cite{dular2019finite}. Also, the \hpf strongly verifies the curl-free condition on $\h$ in $\Occ$, Eq.~\eqref{eq_curlFreeCondition}, by expressing the magnetic field as the gradient of a scalar potential. This leads to a lower number of degrees of freedom compared to the \hfOnly~\cite{Brambilla2006}, that uses instead a spurious
non vanishing resistivity to limit the current density in $\Occ$.


The 3D \hpf reads~\cite{dular2019finite}: from an initial solution at $t=0$, find $\h\in\hsp(\O)$ such that, for $t>0$ and $\forall \h' \in \hspz(\O)$, we have
\begin{align}\label{eq_hform_in_x}
\volInt{\dt(\mu\, \h)}{\h'}{\O} + \volInt{\rho\, \curl \h}{\curl \h'}{\Oc} = \sum_{i\in C_V} \bar V_i \mathcal{I}_i(\h').
\end{align}
The integral over $\O$ of the inner product of $\vec f$ and $\vec g$ is denoted by $\volInt{\vec f}{\vec g}{\O}$, whereas the operator $\mathcal{I}_i(\h)$ gives the circulation of $\h$ around conductor $i$, which is the net current $I_i$ flowing in the conductor. 
The associated voltage is noted $\bar V_i$.  
The function space $\hsp(\O)$ is the subspace of $H(\curlOnly;\O)$ containing functions that are curl-free in $\Occ$ and verify the essential BC and the GC~\cite{dular2023standard}. The space $\hspz(\O)$ is the same space as $\hsp(\O)$ but with homogeneous essential BC and homogeneous GC. For simplicity, we assumed homogeneous natural BC in Eq.~\eqref{eq_hform_in_x}.

\subsection{The \hpf in helicoidal coordinates}

As shown in \cite{milton2006cloaking,nicolet2008geometrical}, in order to express the \hpf Eq.~\eqref{eq_hform_in_x} in helicoidal coordinates, it is sufficient to replace the scalar material parameters $\mu$ and $\rho$ by the tensors $\mutilde$ and $\rhotilde$:
\begin{align}\label{eq_matTensors}
\tilde{\vec \mu} &= \mu\ \mat J^{-1} \mat J^{-\text{T}}\det(\mat J) = \mu \inv T,\\
\tilde{\vec \rho} &= \rho\ \frac{1}{\det(\mat J)}\ \mat J^{\text{T}} \mat J = \rho \mat T,\label{eq_matTensors_2}
\end{align}
with the 
auxiliary tensor $\mat T$, defined by
\begin{align}\label{eq_metric_T}
\mat T = \frac{\mat J^{\text{T}} \mat J}{\det(\mat J)} = \begin{pmatrix}
1 & 0 & -\alpha \xi_2\\
0 & 1 & \alpha \xi_1\\
-\alpha \xi_2 & \alpha \xi_1 & 1 + \alpha^2 (\xi_1^2 + \xi_2^2)
\end{pmatrix},
\end{align}
and its inverse $\inv T$ by
\begin{align}\label{eq_metric_Tinv}
\inv T &= \det(\mat J)\ \mat J^{-1} \mat J^{-\text{T}}\notag \\
&= \begin{pmatrix}
1+\alpha^2\xi_2^2 & \alpha^2 \xi_1 \xi_2 & \alpha \xi_2\\
\alpha^2 \xi_1 \xi_2 & 1+\alpha^2\xi_1^2 & -\alpha \xi_1\\
\alpha \xi_2 & -\alpha \xi_1 & 1
\end{pmatrix}.
\end{align}
This is a consequence of substituting Eqn.~\eqref{eq_oneformTransfo} and \eqref{eq_twoformTransfo} into Eq.~\eqref{eq_hform_in_x} and adding a $\det \mat J$ factor in the volume integral terms. 
Beyond these modifications, all calculations can be performed exactly as in Cartesian coordinates~\cite{nicolet2008geometrical}.

The components of the curl operator in helicoidal coordinates are given by 
\begin{align}\label{eq_curlBlind}
(\curl \h)_{\vec{\xi}} = \begin{pmatrix}
\partial_{\xi_2}h_{\xi_3} - \partial_{\xi_3}h_{\xi_2}\\
\partial_{\xi_3}h_{\xi_1} - \partial_{\xi_1}h_{\xi_3}\\
\partial_{\xi_1}h_{\xi_2} - \partial_{\xi_2}h_{\xi_1}
\end{pmatrix}.
\end{align}
They have the same expression as in Cartesian coordinates, but in terms of the helicoidal coordinates.

\subsection{Conditions for reducing the dimension from 3D to 2D}

In the continuous setting, the problems expressed with Cartesian or helicoidal coordinates are equivalent. Indeed, no approximation is introduced and the change of coordinates is regular. For helicoidally symmetric geometries, there are however clear advantages in working with helicoidal coordinates.

First, it involves integrals over domains with $\xi_3$-independent sections.

Second, $\mat T$ and $\inv T$ are also $\xi_3$-independent, as shown in Eqn.~\eqref{eq_metric_T} and \eqref{eq_metric_Tinv}. As a consequence, both the integrand coefficients and the domains of integration in the weak formulation are $\xi_3$-independent. 

Finally, if the BC on $\Gout$ 
are also $\xi_3$-independent when expressed in helicoidal coordinates, then, the solution $\h$ of the \hpf is $\xi_3$-independent as well. Hence, the integration along the $\xi_3$-direction is trivial and the problem dimension can be reduced from 3D to 2D with a considerable decrease of the computational burden compared to the equivalent 3D problem. 

BC are helicoidally symmetric (HS) in the case of a uniform axial magnetic field excitation. In Section~\ref{sec_helicoidal_symmetric2D}, we describe how the associated 2D problem can be discretized and implemented. We verify the implementation in Section~\ref{sec_filaments_verif3Dmodel}.

By contrast, a transverse magnetic field excitation, i.e., a magnetic field in the $x$-$y$-plane in the Cartesian coordinate system, does not transform into a $\xi_3$-independent field in helicoidal coordinates (see Eq.~\eqref{eq_transverseField}). In this case, the dimension cannot be \textit{directly} reduced from 3D to 2D. However, simplifications are still possible, eventually also leading to a 2D problem in some situations. We present a novel method for such a situation in Sections~\ref{sec_helicoidal_non_symmetric} and \ref{sec_filaments_verifquasi3Dmodel}. This method generalizes the case of HS BC, which just becomes a particular case of the general approach.

\section{Practical Implementation of a Full \hpf \textemdash\ HS-BC}\label{sec_helicoidal_symmetric2D}

Starting from Eq.~\eqref{eq_hform_in_x} with material tensors in Eqn.~\eqref{eq_matTensors}, one could be tempted to implement the \hpf directly as a classical 2D problem with in-plane magnetic field, with the only differences of (i) working in helicoidal coordinates, and (ii) having anisotropic tensors instead of scalar material parameters. But this would not be correct: the fact that the problem is $\xi_3$-independent does not imply that the involved magnetic field has only two non-zero (helicoidal or Cartesian) components.

Due to the full anisotropy of tensors $\mutilde$ and $\rhotilde$, one really has to consider three independent components for the magnetic field $\h$ in the \hpfOnly. To emphasize this, we refer to the resulting formulation as a \textit{full} \hpf in 2D, and we call the associated model the \textit{\tdp\ model}.


In this section, we present a practical implementation of this full \hpfOnly. First, we propose a convenient decomposition of the magnetic field, which allows us to reuse the usual function spaces of classical 2D problems. Then, we discuss the discretization of these function spaces. Finally, we explain how to impose the GC and BC.

\subsection{Decomposition of the magnetic field}

In the \hpfOnly, the magnetic field $\h$ can be decomposed into two parts: an \textit{in-plane} contribution $\hpa$, containing the $\xi_1$ and $\xi_2$-components of $\h$, and an \textit{out-of-plane} contribution $\hpe$, containing only the $\xi_3$-component. We write
\begin{align}\label{eq_h_decomposition}
\h(\xi_1,\xi_2) &= \hpa(\xi_1, \xi_2) + \hpe(\xi_1,\xi_2),
\end{align}
or, explicitly in terms of their helicoidal components,
\begin{align}\label{eq_h_decomposition_components}
\begin{pmatrix}
h_{\xi_1}(\xi_1,\xi_2)\\
h_{\xi_2}(\xi_1,\xi_2)\\
h_{\xi_3}(\xi_1,\xi_2)
\end{pmatrix} &= \begin{pmatrix}
h_{\xi_1}(\xi_1,\xi_2)\\
h_{\xi_2}(\xi_1,\xi_2)\\
0
\end{pmatrix} + \begin{pmatrix}
0\\
0\\
h_{\xi_3}(\xi_1,\xi_2)
\end{pmatrix},
\end{align}
where $\h = \h(\xi_1,\xi_2)$ because the solution is $\xi_3$-independent.
Note that the vectors $\hpa$ and $\hpe$ are not orthogonal.

Because the Jacobian is non-singular, the curl-free condition Eq.~\eqref{eq_curlFreeCondition} reads, in the helicoidal coordinate system:
\begin{align}\label{eq_curl_h_condition}
(\curl \h)_{\vec{\xi}} = \begin{pmatrix}
\partial_{\xi_2}h_{\xi_3}\\
-\partial_{\xi_1}h_{\xi_3}\\
\partial_{\xi_1}h_{\xi_2} - \partial_{\xi_2}h_{\xi_1}
\end{pmatrix} = \vec 0,
\end{align}
from Eq~\eqref{eq_curlBlind} using $\partial_{\xi_3} = 0$. With the decomposition defined in Eq.~\eqref{eq_h_decomposition}, the third component of Eq.~\eqref{eq_curl_h_condition} implies that $\curl \hpa = \vec 0$, which is the same condition as for a classical 2D formulation in which a two-component magnetic field is considered. Then, for the first two components of Eq.~\eqref{eq_curl_h_condition} to be equal to zero, the out-of-plane magnetic field $\hpe$ must be uniform in $\Occ$.

These conditions are introduced in the function space definitions, i.e., they are strongly enforced. They will be made explicit at the space discretization step.

With the explicit decomposition $\h = \hpa + \hpe$, the \hpf reads as follows. From an initial solution at time $t=0$, find $\hpa\in \hsppa(\O)$ and $\hpe \in \hsppe(\O)$ such that, for $t>0$, $\forall \hpa'\in \hspzpa(\O)$ and $\forall\hpe'\in \hspzpe(\O)$,
\begin{align}
&\volInt{\dt(\tilde{\vec \mu}\, (\hpa+\hpe))}{\hpa'}{\O}+ \volInt{\tilde{\vec \rho}\, \curl (\hpa+\hpe)}{\curl \hpa'}{\Oc}\notag\\
&\hspace{4.5cm} = \sum_{i\in C_V} \bar V_i \mathcal{I}_i(\hpa'),\label{eq_hform_parallel}\\
&\volInt{\dt(\tilde{\vec \mu}\, (\hpa+\hpe))}{\hpe'}{\O} + \volInt{\tilde{\vec \rho}\, \curl (\hpa+\hpe)}{\curl \hpe'}{\Oc}\notag \\
&\hspace{4.5cm}  = 0,\label{eq_hform_perpendicular}
\end{align}
where the vectors $\hpa$ and $\hpe$ are coupled by tensors $\mutilde$ and $\rhotilde$. Note that $\mathcal{I}_i(\hpe') = 0$. The function spaces $\hsppa(\O)$ and $\hsppe(\O)$ will be defined in the space discretization step.

For the resistivity in superconducting materials, the power law Eq.~\eqref{eqn_contitutiveje} leads to $\rhotilde = \rho_\text{SC}(\|\j\|) \ \mat T$. Using Eq.~\eqref{eq_twoformTransfo} and $\det \mat J = 1$, we have, in terms of the components: $\|\j\|^2 = \j_{\vec x}\transpose \j_{\vec x} = \j_{\vec \xi}\transpose \mat J\transpose \mat J \j_{\vec \xi} = \j_{\vec \xi}\transpose \mat T \j_{\vec \xi}$, which is $\xi_3$-independent.


\subsection{Space discretization of the magnetic field}

Let us consider a finite element mesh for the discretization of the 2D domain $\O$, and let us denote by $\nodes(\O_i)$ and $\edges(\O_i)$, the set of nodes and edges, respectively, of the mesh in a given (sub-)domain $\O_i$, including entities on the boundary of $\O_i$.

In practice, we can discretize the in-plane magnetic field $\hpa$ exactly as the two-component magnetic field in a classical 2D \hpf with in-plane magnetic field~\cite{dular2019finite}. We use Whitney forms~\cite{bossavit1988whitney}: gradient of node functions $w_n$ and cohomology functions $\vec c_i$ (cut functions)~\cite{pellikka2013homology} in $\Occ$, and edge functions $\vec w_e$ in $\Oc\backslash \partial \Oc$:
\begin{equation}\label{eq_h_parallel_decomposition}
\hpa = \sum_{e \in \edges(\Oc\backslash \partial \Oc)}\hspace{-0.44cm} h_{\parallel, e} \ \vec w_e + \sum_{n \in \nodes(\Occ)}  \hspace{-0.3cm} \phi_n \ \grad w_n + \sum_{i \in C} I_i \ \vec c_i,
\end{equation}
where coefficients $h_{\parallel, e}$, $\phi_n$, and $I_i$ are the degrees of freedom (DOFs) defining $\hpa$ in the discrete function space $\hsppa(\O)$.

We choose to discretize the out-of-plane magnetic field $\hpe$ with perpendicular edge functions $\vec w_n = w_n \ezh$, associated with nodes. 
To account for the fact that $\hpe$ must be uniform in each region of $\Occ$, we introduce global functions in $\Occ$. Let $K$ be the number of connected regions in $\Occ$. We describe the out-of-plane magnetic field with the expansion
\begin{align}\label{eq_hpe_space}
\hpe = \sum_{n\in \nodes(\Oc\setminus \partial \Oc)} h_{\perp,n}\, \vec w_n + \sum_{i=1}^{K} D_i\, \vec p_i,
\end{align}
with $\vec w_n$ the perpendicular edge function associated with node $n$ in $\nodes(\Oc\setminus \partial \Oc)$, and $\vec p_i$ a global shape function defined as the sum of all perpendicular edge functions associated with nodes in the $i^{\text{th}}$ connected region of $\Occ$, including those on its boundary, for $i\in \{1, \dots, K\}$. The support of the shape function $\vec p_i$ is therefore not restricted to $\Occ$: it is non-zero on a layer of one element adjacent to $\partial \Oc$ in $\Oc$. This defines the discrete function space $\hsppe(\O)$, with DOFs $h_{\perp,n}$ and $D_i$. Both $\hpa$ and $\hpe$ are described by discrete \oneforms, and so is their sum, $\h$.

For simplicity, in the following, we assume that there is only one connected non-conducting region $\Occ$, the exterior of the wire, such that $K = 1$, and we rename $D_1 = D$.

\subsection{Global conditions and boundary conditions}

For the GC, a current $\bar I_i$, for $i\in C_I$, can be imposed exactly as in a classical 2D \hpf with in-plane magnetic field, i.e., strongly via the degree of freedom $I_i$ associated with the cut function $\vec c_i$ for the corresponding conducting domain $\Oci$. Alternatively, an applied voltage $\bar V_i$, for $i\in C_V$, can be imposed weakly in the global term of the formulation Eq.~\eqref{eq_hform_parallel}.

For the BC, we consider a circular external boundary $\Gout$, placed in $\Occ$ sufficiently far from the conductors such that we can assume $\dt \b\cdot \n|_{\Gout} = 0$, with $\n$ the outer normal vector. This condition is implicitly imposed for $\hpa$ in Eq.~\eqref{eq_hform_parallel} with homogeneous natural BC on $\Gout$. This lets the $z$-component of the magnetic field, $h_z$, undetermined on $\Gout$. It corresponds to the axial magnetic field, that we can freely impose. We derive below how to translate this into a BC on $\hpe|_{\Gout}$ in helicoidal coordinates.

Let us first consider the situation with a zero axial magnetic field. At a sufficiently large distance $R_\text{out}$ from the center of conductors carrying a total net current intensity $I$, the magnetic field tends to be purely azimuthal and axisymmetric. We have $\h_{\vec x} = \frac{I}{2\pi R_\text{out}} (-\sin \theta \quad \cos \theta \quad 0)^{\text{T}}$, with $\theta = \atantwo(y,x)$. In terms of the helicoidal coordinates, on the plane $\xi_3 = 0$, it reads
\begin{align}
\h_{\vec \xi} =  \mat J^{\text{T}}|_{\xi_3 = 0}\ \h_{\vec x} &= \frac{I}{2\pi R_\text{out}} \begin{pmatrix}
-\sin \theta\\
\cos \theta\\
\alpha \xi_2 \sin \theta + \alpha \xi_1 \cos \theta
\end{pmatrix}\notag\\
& =  
\frac{I}{2\pi R_\text{out}} \begin{pmatrix}
-\sin \theta\\
\cos \theta\\
\alpha R_\text{out}
\end{pmatrix},
\end{align}
using $\xi_2 = R_\text{out} \sin \theta$ and $\xi_1 = R_\text{out}\cos \theta$ for $\xi_3=0$. Consequently, to satisfy $h_z|_{\Gout} = 0$, one has to impose that $h_{\xi_3}|_{\Gout} = I\alpha/2\pi$. This can be done by fixing the degree of freedom $D$ associated with the basis function $\vec p$ in $\Occ$ in Eq.~\eqref{eq_hpe_space} to the value $D = I\alpha/2\pi$. Note that this value does not depend on $R_\text{out}$.

By superposition, if one wants to impose a non-zero axial magnetic field $h_\text{axial}$ on the external boundary $\Gout$ in addition to a net current intensity $I$, we can impose the following condition:
\begin{align}\label{eq_hperp_condition}
D = \frac{I\alpha}{2\pi} + h_\text{axial},
\end{align}
because the axial magnetic field Cartesian components $\h_{\vec x} = (0 \quad 0 \quad h_\text{axial})^{\text{T}}$ transform into $\h_{\vec \xi} =  \mat J^{\text{T}}\ \h_{\vec x} = (0 \quad 0 \quad h_\text{axial})^{\text{T}}$ in helicoidal coordinates.

\section{Verification and Application \textemdash\ HS-BC}\label{sec_filaments_verif3Dmodel}

In this section, we first compare the solution of the \tdp\  model in helicoidal coordinates to the solution of a classical 3D \hpf on a simple problem in order to verify the implementation. We also quantify the computational gain offered by reducing the dimension from 3D to 2D. Then, we apply the \tdp\  model on a more involved geometry to illustrate the capabilities of the approach.

\subsection{Verification problem}

\begin{figure}[h!]
      \begin{subfigure}[b]{0.49\linewidth}  
            \centering 
		\includegraphics[width=0.85\textwidth]{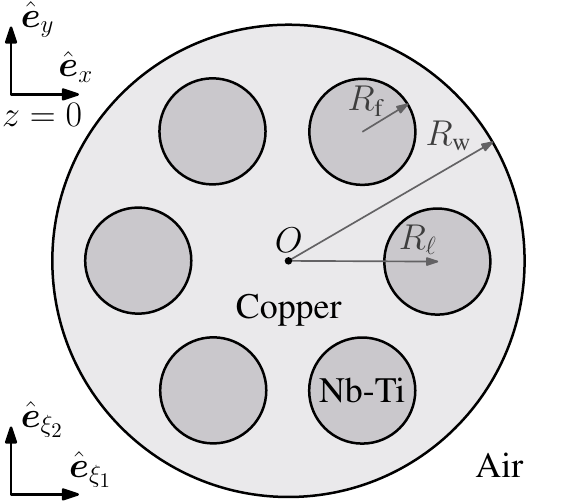}
		\caption{Wire geometry in 2D.}
		\label{filamentGeometry_2D}	
      \end{subfigure}
      \begin{subfigure}[b]{0.49\linewidth}  
            \centering 
		\includegraphics[width=\textwidth]{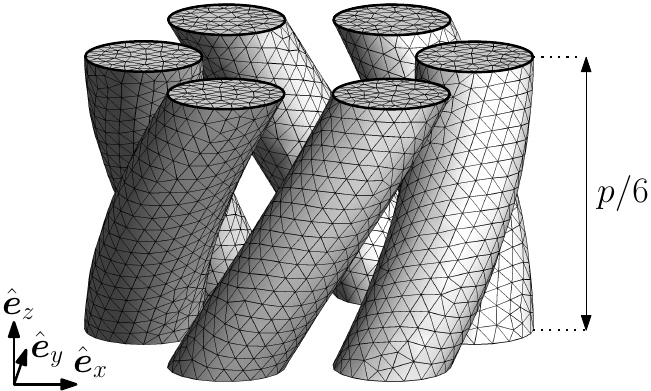}
		\caption{Wire geometry in 3D.}
		\label{filamentGeometry}	
      \end{subfigure}
\caption{Wire geometry for the verification of the helicoidal transformation consisting of six twisted Nb-Ti filaments embedded in a copper matrix. (a) Geometry in a $\xi_1$-$\xi_2$ plane (or in the $x$-$y$ plane for $z=0$). (b) One-sixth of a pitch length. The copper matrix in (b) is not represented, for clarity.}
\label{filamentGeometry_full}
\end{figure}

We consider a wire made of six identical Nb-Ti superconducting filaments, twisted and embedded in a copper (Cu) matrix, as illustrated in Fig.~\ref{filamentGeometry_full}. In order to simplify the geometry, the cross sections of the filaments are assumed to be disks, and the 3D geometry is generated by a helicoidal extrusion of them. This is of course an approximation of a realistic geometry. If needed, cross sections of round twisted filaments can be computed accurately using envelope theory as in~\cite{piwonski20232d} or CAD tools~\cite{gmsh} as in~\cite{piwonski2023_aform}.

The filaments have a radius of $R_\text{f} = 35$ $\mu$m and their centers are at a distance $R_\ell = 98$ $\mu$m from the center of the wire. The wire has a radius of $R_\text{w}=155$ $\mu$m and a pitch length of $p=1$~mm. The air is modelled outside of the wire up to a distance $R_\text{out} = 500$ $\mu$m.

We assume that Nb-Ti resistivity is characterized by Eq.~\eqref{eqn_contitutiveje} with constant and uniform $\jc=7\times 10^9$~A/m$^2$ and $n=50$, and that the copper resistivity is $\rho_\text{Cu} = 1.81 \times 10^{-10}$~$\Omega$m. There is no insulation between the filaments and the matrix, so that the wire behaves as a single conducting cylinder. A net transport current $I(t) = 0.5\, I_\text{c} \sin(2\pi t/T)$ is imposed in the wire, with $T = 0.1$~s and $I_\text{c} = 162$~A, and we impose $h_\text{axial} =0$~A/m.

\subsection{Implementation of the 3D model}

\begin{figure}[h!]
      \centering
      \includegraphics[width=0.85\linewidth]{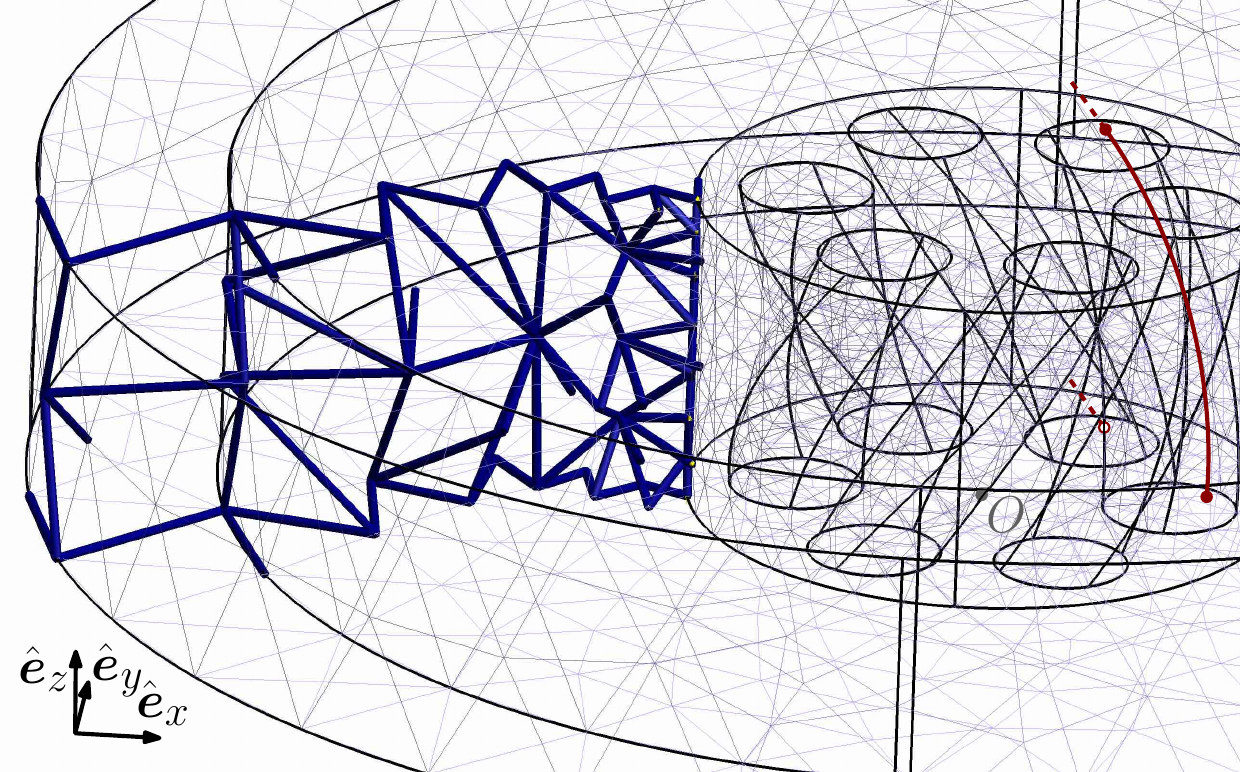}
\caption{Periodic support for the cohomology basis function to impose a transport current $I(t)$ in the 3D verification model. The red curve in the filaments is a portion of the helicoidal fiber along which the solution is represented in Fig.~\ref{filament_verif_helicoidal_3D_fine}.}
\label{3Dcut_condMatrix}
\end{figure}

We consider the 3D geometry represented in Fig.~\ref{filamentGeometry}. It represents a periodic cell of one-sixth of a whole pitch length $p$. Note that building and meshing the 3D model represented in Fig.~\ref{filamentGeometry} is not a trivial task. To account for the periodicity of the problem, the mesh must be identical on the top and bottom boundaries of the domain; as an \hpf is used, cohomology basis functions must also be periodic. The quality of the mesh inside the filaments plays an important role for the accuracy of the resulting numerical solution. We observed that better results are obtained with a structured mesh inside the filaments. Generating the mesh with such constraints is possible with Gmsh~\cite{gmsh}. The periodic support for the cohomology basis function is generated as described in~\cite{pellikka2013homology,desousa20213} and illustrated in Fig.~\ref{3Dcut_condMatrix}.

We set a homogeneous natural BC on the external boundary $\Gout$, so that $\dt \b\cdot \n|_{\Gout} = 0$ is weakly enforced. For the top and bottom boundaries $\Gamma_\text{up}$ and $\Gamma_\text{down}$, which are topologically identical, the periodic condition $\h\times \n|_{\Gamma_\text{up}} = - \h\times \n|_{\Gamma_\text{down}}$ is imposed. On conducting boundaries $\partial \Oc \cap (\Gamma_\text{up} \cup \Gamma_\text{down})$, this is done by forcing the equality of the degrees of freedom associated with topologically identical edges of these boundaries. On non-conducting boundaries $\partial \Occ \cap (\Gamma_\text{up} \cup \Gamma_\text{down})$, the periodic constraint is enforced via the magnetic scalar potential. We impose $\phi|_{\Gamma_\text{up}} = \phi|_{\Gamma_\text{down}} + h_\text{axial}\, p/6$. The total current intensity flowing in the conducting domain made up of the filaments and matrix is imposed via the (periodic) cohomology basis function whose generating edges are highlighted in Fig.~\ref{3Dcut_condMatrix}.

Note that in the present case of HS-BC (transport current or axial field), the 3D reference model could be defined on a length shorter than $p/6$ along $z$, if one adapts the periodic mesh and the periodic cut accordingly. We chose a length of $p/6$ so that the reference model will also be valid in the transverse field case.

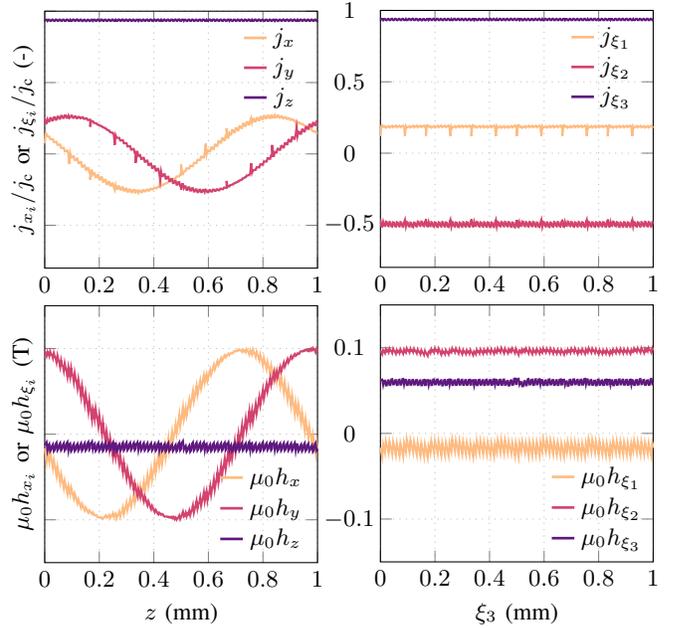
\begin{figure}[h!]
        \centering
        \begin{subfigure}[b]{0.49\linewidth}
        \centering
             \tikzsetnextfilename{filament_verif_helicoidal_3D_x_fine}
	\begin{tikzpicture}[trim axis left, trim axis right][font=\small]
\begin{groupplot}[group style={group size=1 by 2,
       horizontal sep=0pt,
       vertical sep=0.5cm},
     ] 
 	\nextgroupplot[
    width=1.2\linewidth,
    height=5cm,
    grid = both,
    grid style = dotted,
    xmin=0, 
    xmax=1,
    ymin=-0.8, 
    ymax=1,
    yticklabel=\empty,
	xtick scale label code/.code={},
	ytick scale label code/.code={},
    ylabel={$j_{x_i}/j_\text{c}$ or $j_{\xi_i}/j_\text{c}$ (-)},
    ylabel style={yshift=-3em},
    xlabel style={yshift=0.2em},
    legend style={at={(0.97, 0.78)}, cells={anchor=east}, anchor= east, draw=none, fill opacity=0, text opacity = 1,legend image code/.code={\draw[##1,line width=1pt] plot coordinates {(0cm,0cm) (0.3cm,0cm)};}}
    ]
    \addplot[mag_6, thick] 
    table[x=z,y=jx]{data/filament_verif_helicoidal_3D_fine.txt};
    \addplot[mag_4, thick] 
    table[x=z,y=jy]{data/filament_verif_helicoidal_3D_fine.txt};
        \addplot[mag_2, thick] 
    table[x=z,y=jz]{data/filament_verif_helicoidal_3D_fine.txt};
    \legend{$j_x$, $j_y$, $j_z$}
\nextgroupplot[
    width=1.2\textwidth,
    height=5cm,
    grid = both,
    grid style = dotted,
    xmin=0, 
    xmax=1,
    ymin=-0.15, 
    ymax=0.15,
    yticklabel=\empty,
	xtick scale label code/.code={},
    yticklabel style={anchor=center},
	ytick scale label code/.code={},
	xlabel={$z$ (mm)},
    ylabel={$\mu_0 h_{x_i}$ or $\mu_0 h_{\xi_i}$ (T)},
    ylabel style={yshift=-3em},
    xlabel style={yshift=0.3em},
    legend style={at={(0.8, 0.01)}, anchor=south, draw=none, fill opacity=0, text opacity = 1,legend image code/.code={\draw[##1,line width=1pt] plot coordinates {(0cm,0cm) (0.3cm,0cm)};}}
    ]
    \addplot[mag_6, thick] 
    table[x=z,y=bx]{data/filament_verif_helicoidal_3D_fine.txt};
    \addplot[mag_4, thick] 
    table[x=z,y=by]{data/filament_verif_helicoidal_3D_fine.txt};
        \addplot[mag_2, thick] 
    table[x=z,y=bz]{data/filament_verif_helicoidal_3D_fine.txt};
    \legend{$\mu_0 h_x$, $\mu_0 h_y$, $\mu_0 h_z$}
    \end{groupplot}
	\end{tikzpicture}%
        \label{filament_verif_helicoidal_3D_x_fine}
	        \end{subfigure}
        \begin{subfigure}[b]{0.49\linewidth}  
\centering
             \tikzsetnextfilename{filament_verif_helicoidal_3D_xi_fine}
	\begin{tikzpicture}[trim axis left, trim axis right][font=\small]
\begin{groupplot}[group style={group size=1 by 2,
       horizontal sep=0pt,
       vertical sep=0.5cm},
     ] 
  	\nextgroupplot[
    width=1.2\linewidth,
    height=5cm,
    grid = both,
    grid style = dotted,
    xmin=0, 
    xmax=1,
    ymin=-0.8, 
    ymax=1,
	xtick scale label code/.code={},
	ytick scale label code/.code={},
    ylabel style={yshift=-0.5em},
    xlabel style={yshift=0.2em},
    yticklabel style={anchor=center},
    yticklabel style={xshift=-1.3em},
    legend style={at={(0.97, 0.78)}, cells={anchor=east}, anchor= east, draw=none, fill opacity=0, text opacity = 1,legend image code/.code={\draw[##1,line width=1pt] plot coordinates {(0cm,0cm) (0.3cm,0cm)};}}
    ]
    \addplot[mag_6, thick] 
    table[x=z,y=jxi1]{data/filament_verif_helicoidal_3D_fine.txt};
    \addplot[mag_4, thick] 
    table[x=z,y=jxi2]{data/filament_verif_helicoidal_3D_fine.txt};
        \addplot[mag_2, thick] 
    table[x=z,y=jxi3]{data/filament_verif_helicoidal_3D_fine.txt};
    \legend{$j_{\xi_1}$, $j_{\xi_2}$, $j_{\xi_3}$}
\nextgroupplot[
    width=1.2\textwidth,
    height=5cm,
    grid = both,
    grid style = dotted,
    xmin=0, 
    xmax=1,
    ymin=-0.15, 
    ymax=0.15,
	xtick scale label code/.code={},
    yticklabel style={anchor=center},
	ytick scale label code/.code={},
	xlabel={$\xi_3$ (mm)},
    ylabel style={yshift=-0.5em},
    xlabel style={yshift=0.3em},
    yticklabel style={anchor=center},
    yticklabel style={xshift=-1.3em},
    legend style={at={(0.8, 0.01)}, anchor=south, draw=none, fill opacity=0, text opacity = 1,legend image code/.code={\draw[##1,line width=1pt] plot coordinates {(0cm,0cm) (0.3cm,0cm)};}}
    ]
    \addplot[mag_6, thick] 
    table[x=z,y=bxi1]{data/filament_verif_helicoidal_3D_fine.txt};
    \addplot[mag_4, thick] 
    table[x=z,y=bxi2]{data/filament_verif_helicoidal_3D_fine.txt};
    \addplot[mag_2, thick] 
    table[x=z,y=bxi3]{data/filament_verif_helicoidal_3D_fine.txt};
    \legend{$\mu_0 h_{\xi_1}$, $\mu_0 h_{\xi_2}$, $\mu_0 h_{\xi_3}$}
    \end{groupplot}
	\end{tikzpicture}%
        \label{filament_verif_helicoidal_3D_xi_fine}
            \end{subfigure}
        \caption{Current density (up) and magnetic field (down) components along a helicoidal fiber from $z=0$ to $z=p$. (Left) Cartesian components of the vectors. (Right) Helicoidal components of the vectors. Solution at $t=T/4$.}
        \label{filament_verif_helicoidal_3D_fine}
\end{figure}

Before comparing the results, we first verify that the 3D model indeed produces a helicoidally symmetric solution. For illustration, from the 3D numerical solution, we extract the magnetic field $\h$ and the current density $\j$ along the helicoidal fiber of pitch length $p$ passing at point $\vec x = \big(a,b, 0\big)$, with $a=180$~mm and $b=11$~mm, from $z=0$ to $z=p$ (see Fig.~\ref{3Dcut_condMatrix}). We exploit the periodicity of the problem to obtain values for $z>p/6$. The Cartesian and helicoidal components of vectors $\h$ and $\j$ are represented in Fig.~\ref{filament_verif_helicoidal_3D_fine} for a relatively fine tetrahedral mesh ($144\,870$ DOFs), at time $t=T/4$. Helicoidal components are obtained using the one and two-forms transformation relations, Eqn.~\eqref{eq_oneformTransfo}~and~\eqref{eq_twoformTransfo}.  

The oscillations and spikes along the fiber represent inter-element 
non-conformities, which are expected with 
lowest-order tetrahedral Whitney shape functions. 
These oscillations decrease in amplitude with mesh refinement. Up to these inter-element variations, the 3D solution correctly presents a helicoidal symmetry. It is also interesting to notice that the current density has non-zero $\xi_1$ and $\xi_2$-components, and that the $\xi_3$-component of the magnetic field is not equal to zero. This illustrates the need for a three-component magnetic field in the 2D helicoidal model.

\subsection{Comparison of the results from the 3D and \tdp\ models}

We now compare the results of the \tdp\ problem in helicoidal coordinates with the reference 3D problem described above. Note that for the \tdp\ model, in this particular case, we could further exploit the symmetry and model only one-sixth of the circular region depicted in Fig.~\ref{filamentGeometry_2D} using periodic BC on the symmetry boundaries as well as an adapted cohomology function in $\Occ$, hence reducing the computational cost even more. We however choose to model the full 2D cross section.

\begin{figure}[h!]
\begin{center}
\includegraphics[width=\linewidth]{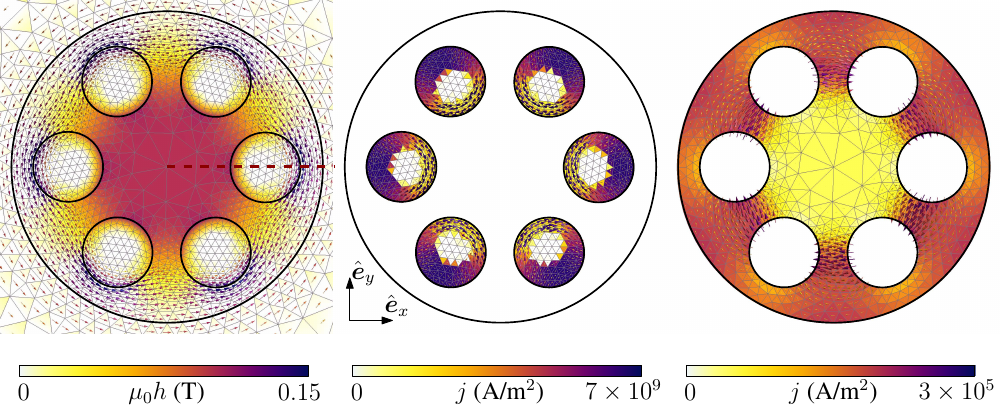}
\caption{Magnetic field (left), current density in the filaments (middle), and current density in the matrix (right) at time $t=T/4$ for the \tdp\ problem solved with the helicoidal coordinate system. The arrows represent the in-plane $x$ and $y$-components of $\h$ and $\j$, whereas the triangular elements are colored as a function of the out-of-plane $z$-component of $\h$ and $\j$. The dashed red line in the left figure is the cut along which the magnetic field is represented in Fig.~\ref{verifB_super_fine}.}
\label{filament_b_and_j}
\end{center}
\end{figure}

The solution of the \tdp\ model on a medium mesh resolution ($4\,700$ DOFs) is represented in Fig.~\ref{filament_b_and_j}. The current mostly flows in the superconducting filaments, as shown by the different scales for the middle and right subfigures. On the left subfigure, one can see that the current flow in the twisted filaments induces a non-zero $z$-component $h_z$ of the magnetic field at the center of the wire.

\begin{figure}[h!]
 \begin{subfigure}[b]{0.99\linewidth}  
        \centering
             \tikzsetnextfilename{verifB_super2}
\begin{tikzpicture}[trim axis left, trim axis right][font=\small]
 	\begin{axis}[
    width=\linewidth,
    height=5cm,
    grid = both,
    grid style = dotted,
    ymin=-0.04, 
    ymax=0.2,
    xmin=0, 
    xmax=380,
     yticklabel style={xshift=-0em},
    yticklabel pos=right,
    xticklabel style={yshift=-0em},
    ylabel={$\mu_0 h_{x}$ or $\mu_0 h_{y}$ (T)},
    ylabel style={yshift=-2.5em},
    xlabel style={yshift=0.2em},
    legend style={at={(0.99, 0.99)}, cells={anchor=west}, anchor=north east, draw=none,fill opacity=0, text opacity = 1, legend image code/.code={\draw[##1,line width=1pt] plot coordinates {(0cm,0cm) (0.3cm,0cm)};}}
    ]
    \addplot[mag_4, thick, densely dashed] 
    table[x=r,y=by_3D]{data/verifB_super2.txt};
    \addplot[mag_2, thick, densely dashed]
    table[x=r,y=bz_3D]{data/verifB_super2.txt};
    \addplot[mag_4, thick] 
    table[x=r,y=by_2D]{data/verifB_super2.txt};
        \addplot[mag_2, thick] 
    table[x=r,y=bz_2D]{data/verifB_super2.txt};
     \addplot[gray] 
    coordinates {(63,-1) (63, 1)}; 
         \addplot[gray] 
    coordinates {(133,-1) (133, 1)};
    \addplot[gray] 
     coordinates {(155,-1) (155, 1)}; 
     \node[] at (axis cs: 31.5,0.18) {Cu};
     \node[] at (axis cs: 98,0.18) {Nb-Ti};
    	 \node[] at (axis cs: 144,0.18) {Cu};
     \node[] at (axis cs: 180,0.18) {Air};
    \legend{3D - $\mu_0 h_y$ (coarse), 3D - $\mu_0 h_z$ (coarse), \tdp\ - $\mu_0 h_y$ (coarse), \tdp\ - $\mu_0 h_z$ (coarse)}
    \end{axis}
\end{tikzpicture}%
	\end{subfigure}
        \hfill\vspace{-0.2cm}
        \begin{subfigure}[b]{0.988\linewidth}  
        \centering
             \tikzsetnextfilename{verifB_super_fine}
\begin{tikzpicture}[trim axis left, trim axis right][font=\small]
 	\begin{axis}[
    width=\linewidth,
    height=5cm,
    grid = both,
    grid style = dotted,
    ymin=-0.04, 
    ymax=0.2,
    xmin=0, 
    xmax=380,
     yticklabel style={xshift=-0em},
    yticklabel pos=right,
    xticklabel style={yshift=-0em},
	xlabel={Position $x$ ($\mu$m)},
    ylabel={$\mu_0 h_{x}$ or $\mu_0 h_{y}$ (T)},
    ylabel style={yshift=-2.5em},
    xlabel style={yshift=0.5em},
    legend style={at={(0.99, 0.99)}, cells={anchor=west}, anchor=north east, draw=none,fill opacity=0, text opacity = 1, legend image code/.code={\draw[##1,line width=1pt] plot coordinates {(0cm,0cm) (0.3cm,0cm)};}}
    ]
    \addplot[mag_4, thick, densely dashed] 
    table[x=r,y=by_3D]{data/verifB_super_fine.txt};
    \addplot[mag_2, thick, densely dashed]
    table[x=r,y=bz_3D]{data/verifB_super_fine.txt};
    \addplot[mag_4, thick] 
    table[x=r,y=by_2D]{data/verifB_super_fine.txt};
        \addplot[mag_2, thick] 
    table[x=r,y=bz_2D]{data/verifB_super_fine.txt};
     \addplot[gray] 
    coordinates {(63,-1) (63, 1)}; 
         \addplot[gray] 
    coordinates {(133,-1) (133, 1)};
    \addplot[gray] 
     coordinates {(155,-1) (155, 1)}; 
     \node[] at (axis cs: 31.5,0.18) {Cu};
     \node[] at (axis cs: 98,0.18) {Nb-Ti};
    	 \node[] at (axis cs: 144,0.18) {Cu};
     \node[] at (axis cs: 180,0.18) {Air};
    \legend{3D - $\mu_0 h_y$ (fine), 3D - $\mu_0 h_z$ (fine), \tdp\ - $\mu_0 h_y$ (fine), \tdp\ - $\mu_0 h_z$ (fine)}
    \end{axis}
\end{tikzpicture}%
\end{subfigure}
 \hfill
\vspace{-0.2cm}
\caption{Magnetic field along the dashed red line represented in Fig.~\ref{filament_b_and_j}, for the 3D and \tdp\ models, at time $t=T/4$, with coarse (up) and fine (down) mesh resolutions.}
        \label{verifB_super_fine}
\end{figure}
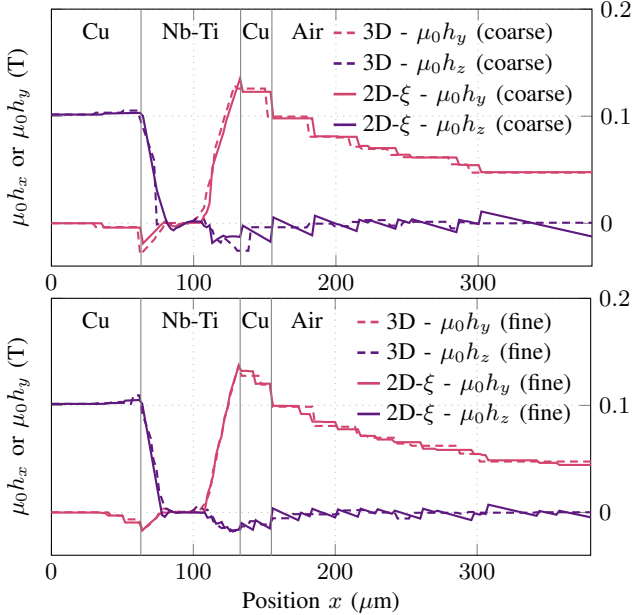

A comparison of the local magnetic field of the \tdp\ model with that of the reference 3D model is given in Fig.~\ref{verifB_super_fine}, along the dashed red line highlighted in Fig.~\ref{filament_b_and_j}, for two mesh resolutions. The solution of the 3D model is taken on the plane $z = \xi_3 = 0$, but this choice is arbitrary: as was shown in Fig.~\ref{filament_verif_helicoidal_3D_fine}, up to the inter-element variation, the solution of the 3D model is also $\xi_3$-independent. Solutions of the \tdp\ and 3D models match locally. We verified and this is also the case for the current density (not represented in the figures).

\begin{figure}[h!]
        \centering
                \begin{subfigure}[b]{0.99\linewidth}  
\centering
             \tikzsetnextfilename{filament_AC_losses}
\begin{tikzpicture}[trim axis left, trim axis right][font=\small]
\pgfplotsset{set layers}
 	\begin{axis}[
	tick label style={/pgf/number format/fixed},
    width=\linewidth,
    height=5cm,
    grid = both,
    grid style = dotted,
    ymin=0, 
    ymax=7.5,
    xmin=0, 
    xmax=0.1,
    ylabel={AC loss (mW/m)},
    ylabel style={yshift=-2.5em},
    ylabel style={xshift=0em},
    xlabel style={yshift=0.4em},
    xticklabel style={yshift=-0.4em},
    yticklabel style={xshift=0em},
    yticklabel pos=right,
    legend style={at={(0.33, 1)}, cells={anchor=west}, anchor=north, draw=none,fill opacity=0, text opacity = 1}
    ]
    \addplot[mag_6, thick, densely dashed] 
    table[x=t,y=hts_3D_coarse]{data/filament_AC_losses.txt};
    \addplot[mag_2, thick, densely dashed] 
    table[x=t,y=hts_2D_coarse]{data/filament_AC_losses.txt};
    \addplot[mag_6, thick] 
    table[x=t,y=hts_3D_fine]{data/filament_AC_losses_fine.txt};
    \addplot[mag_2, thick] 
    table[x=t,y=hts_2D_fine]{data/filament_AC_losses.txt};
    \legend{3D - Nb-Ti (coarse), \tdp\ - Nb-Ti (coarse), 3D - Nb-Ti (fine), \tdp\ - Nb-Ti (fine)}
    \end{axis}
\end{tikzpicture}%
	\end{subfigure}
        \hfill
        \begin{subfigure}[b]{0.99\linewidth}  
\centering
             \tikzsetnextfilename{filament_AC_losses_matrix}
\begin{tikzpicture}[trim axis left, trim axis right][font=\small]
\pgfplotsset{set layers}
 	\begin{axis}[
	tick label style={/pgf/number format/fixed},
    width=\linewidth,
    height=4.5cm,
    grid = both,
    grid style = dotted,
    ymin=0, 
    ymax=0.15,
    xmin=0, 
    xmax=0.1,
	xlabel={Time (s)},
    ylabel={AC loss (mW/m)},
    ylabel style={yshift=-2.5em},
    xlabel style={yshift=0.4em},
    xticklabel style={yshift=-0.4em},
    yticklabel style={xshift=0em},
    yticklabel pos=right,
    legend style={at={(0.27, 0.995)}, cells={anchor=west}, anchor=north, draw=none,fill opacity=0, text opacity = 1}
    ]
    \addplot[mag_6, thick, densely dashed] 
    table[x=t,y=copper_3D_coarse]{data/filament_AC_losses.txt};
    \addplot[mag_2, thick, densely dashed] 
    table[x=t,y=copper_2D_coarse]{data/filament_AC_losses.txt};
    \addplot[mag_6, thick] 
    table[x=t,y=copper_3D_fine]{data/filament_AC_losses_fine.txt};
    \addplot[mag_2, thick] 
    table[x=t,y=copper_2D_fine]{data/filament_AC_losses.txt};
    \legend{3D - Cu (coarse), \tdp\ - Cu  (coarse), 3D - Cu (fine), \tdp\ - Cu (fine)}
    \end{axis}
\end{tikzpicture}%
\end{subfigure}
\vspace{-0.2cm}
\caption{AC losses in the superconducting filaments (up) and in the conducting matrix (down) for a transport current $I(t)$, as a function of time, for two mesh resolutions, with the \tdp\ model in helicoidal coordinates and the 3D verification model.}
        \label{filament_AC_losses}
\end{figure}
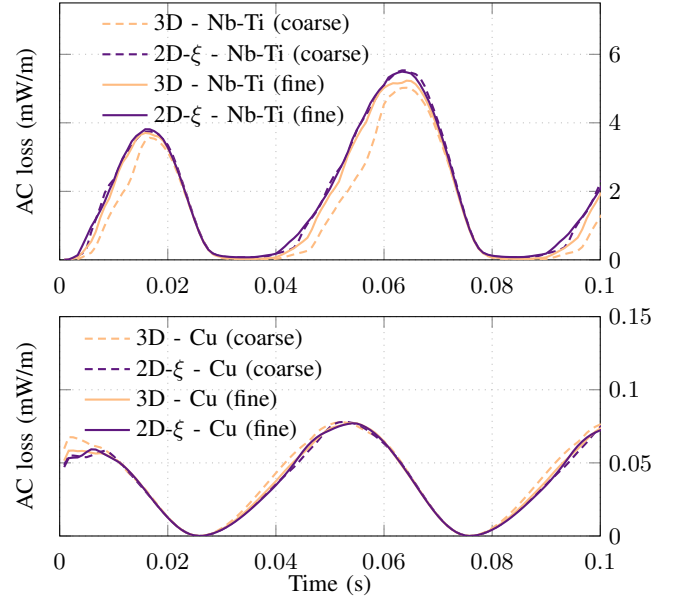

A comparison of the AC loss is given in Fig.~\ref{filament_AC_losses}. The AC loss per unit length along $\ez$ in both the superconducting filaments and the conducting matrix are compared for the two models, and for two mesh resolutions. For the \tdp\ model in helicoidal coordinates, the AC loss is computed as $\volInt{\rhotilde\, \j_{\vec \xi}}{\j_{\vec \xi}}{\Oc}$, where $\Oc$ is either restricted to the filaments, or to the matrix. For the 3D model, the integral $\volInt{\rho\, \j_{\vec x}}{\j_{\vec x}}{\Oc}$ is computed over the 3D domain with Cartesian coordinate system, and the result is divided by $p/6$, to obtain the AC loss per unit length as well. Note that both models include all loss contributions by construction: hysteresis losses in the filaments, as well as coupling and eddy current losses in the matrix~\cite{campbell1982general}.

Meshes for the coarse resolution in the $z=\xi_3=0$ plane are similar for the \tdp\ and 3D models, as well as meshes for the fine resolution. However, we observed that the \tdp\ model solution is less sensitive to the mesh resolution. This is due to the inter-element non-conformities in a tetrahedral 3D mesh, which should be made significantly lower (by mesh refinement) to ensure an accurate evaluation of the quadratic quantity representing the AC losses. Meshes with prisms, i.e., extruded triangles, in the filaments were also tested. They give slightly better results, but also increase the complexity of the meshing step, as pyramids must be used as transition elements between prisms in the filaments and tetrahedra outside of them.

The local and global quantity agreement shows the validity of the \tdp\ model in helicoidal coordinates. The dimension reduction allows for a very large reduction of the computational cost. This is demonstrated in Table~\ref{filament_efficiency}, which compares the performance of the \tdp\ and 3D models on meshes with similar characteristic length for the finite elements (triangles in \tdp\ and tetrahedra in 3D). The fine \tdp\ model is more than two orders of magnitude faster to solve than the fine 3D model.

\begin{table}[!h]\small
\centering
\begin{tabular}{l l r c r r}
\hline
\multicolumn{2}{c}{Model \& mesh} & \# DOFs & \# it. & Time/it./DOF & Total time\\
\hline
\multirow{3}{*}{3D} & Coarse & 16\,556 & 1\,645 & 110 $\upmu$s\quad~ & 50 m\\
 & Medium & 85\,605 & 2\,893 & 260 $\upmu$s\quad~ & 17 h 53 m\\
 & Fine & 144\,870 & 3\,255 & 315 $\upmu$s\quad~ & 41 h 23 m\\
 \hline
 \multirow{3}{*}{\tdp} & Coarse & 1\,797 & 1\,299 & 72 $\upmu$s\quad~ & 3 m\\
 & Medium & 4\,481 & 1\,948 & 71 $\upmu$s\quad~ & 10 m\\
 & Fine & 6\,002 & 2\,258 & 76 $\upmu$s\quad~ & 17 m\\
 \hline
\end{tabular}
\caption{Performance comparison for the 3D and \tdp\ models with imposed current and no axial magnetic field, computed with 150 time steps from $t=0$ to $t=5\,T/4$ on a single Intel Core i7 2.2 GHz CPU. DOF: degree of freedom. It.: iteration (Newton-Raphson).}
\label{filament_efficiency}
\end{table}

\subsection{Application to a 54-filament wire}\label{sec_filaments_variousApplications}

\begin{figure}[h!]
\begin{center}
\includegraphics[width=0.5\linewidth]{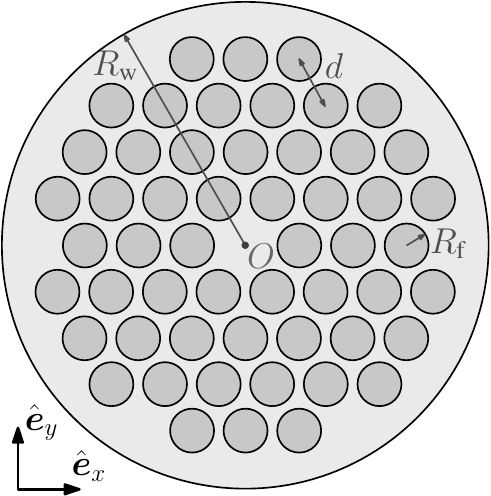}
\caption{Cross section of the 54-filament Nb-Ti/Cu wire.}
\label{54fil_appCurrent}
\end{center}
\end{figure}

As a more realistic geometry, we consider a wire with 54 filaments arranged in a hexagonal lattice with filament center spacing of $d = 110$ $\mu$m, as represented in Fig.~\ref{54fil_appCurrent}. Filament radius is $R_\text{f} = 45$ $\mu$m, wire radius is $R_\text{w} = 500$ $\mu$m, and the pitch length is $p=10$ mm. We keep the same material parameters as before for the Nb-Ti and the copper matrix and we impose a transport current $I(t) = 0.8\, I_\text{c} \sin(2\pi t/T)$, with $T = 1$ s and $I_\text{c} = 2.4$ kA.

\begin{figure}[h!]
\centering
\tikzsetnextfilename{fil_54_current}
\begin{tikzpicture}[trim axis left, trim axis right][font=\small]
  \begin{axis}[
  	tick scale binop=\times,
    width=1\linewidth,
    height=5cm,
    grid = both,
    grid style = dotted,
    xmin=0.1,  
    xmax=0.4,
    xlabel={Distance to the center ($\mu$m)},
    xtick={0.11, 0.22, 0.33},
    xticklabels={110, 220, 330},
    yticklabel pos=right,
    ylabel={Current per filament (A)},
    ylabel style={yshift=-2.5em},
    legend columns=3,
    transpose legend,
    legend style={at={(0.01, 0.01)}, cells={anchor=west}, anchor=south west, draw=none, fill opacity=0, text opacity = 1,style={column sep=0.22cm}, inner sep=1.8pt}
    ]
    \addplot[mag_6, thick, mark=*, mark options={mag_6, scale=0.5, style={solid}}] 
    table[x=r,y=I_50]{data/I_54fil_appCurrent.txt};
        \addplot[mag_5, thick, mark=*, mark options={mag_5, scale=0.5, style={solid}}] 
    table[x=r,y=I_100]{data/I_54fil_appCurrent.txt};
        \addplot[mag_4, thick, mark=*, mark options={mag_4, scale=0.5, style={solid}}] 
    table[x=r,y=I_150]{data/I_54fil_appCurrent.txt};
        \addplot[mag_3, thick, mark=*, mark options={mag_3, scale=0.5, style={solid}}] 
    table[x=r,y=I_200]{data/I_54fil_appCurrent.txt};
        \addplot[mag_2, thick, mark=*, mark options={mag_2, scale=0.5, style={solid}}] 
    table[x=r,y=I_250]{data/I_54fil_appCurrent.txt};
        \addplot[black, thick, mark=*, mark options={mag_1, scale=0.5, style={solid}}] 
    table[x=r,y=I_500]{data/I_54fil_appCurrent.txt};
    \legend{$t=50$ ms, $t=100$ ms, $t=150$ ms, $t=200$ ms, $t=250$ ms, $t=500$ ms}
    \end{axis}
\end{tikzpicture}%
\caption{Distribution of the current among the filaments, as a function of their distance to the center $O$, for the 54-filament geometry.}
\label{fil_54_current}
\end{figure}
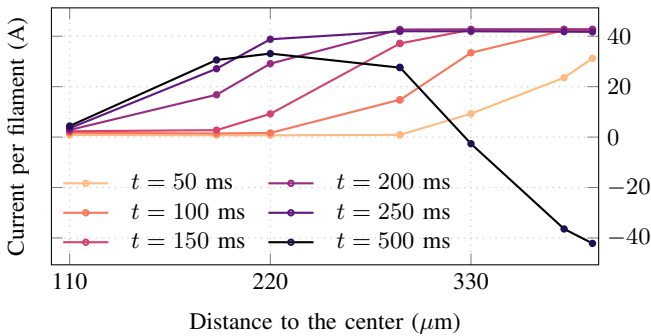

\begin{figure}[h!]
\begin{center}
\includegraphics[width=\linewidth]{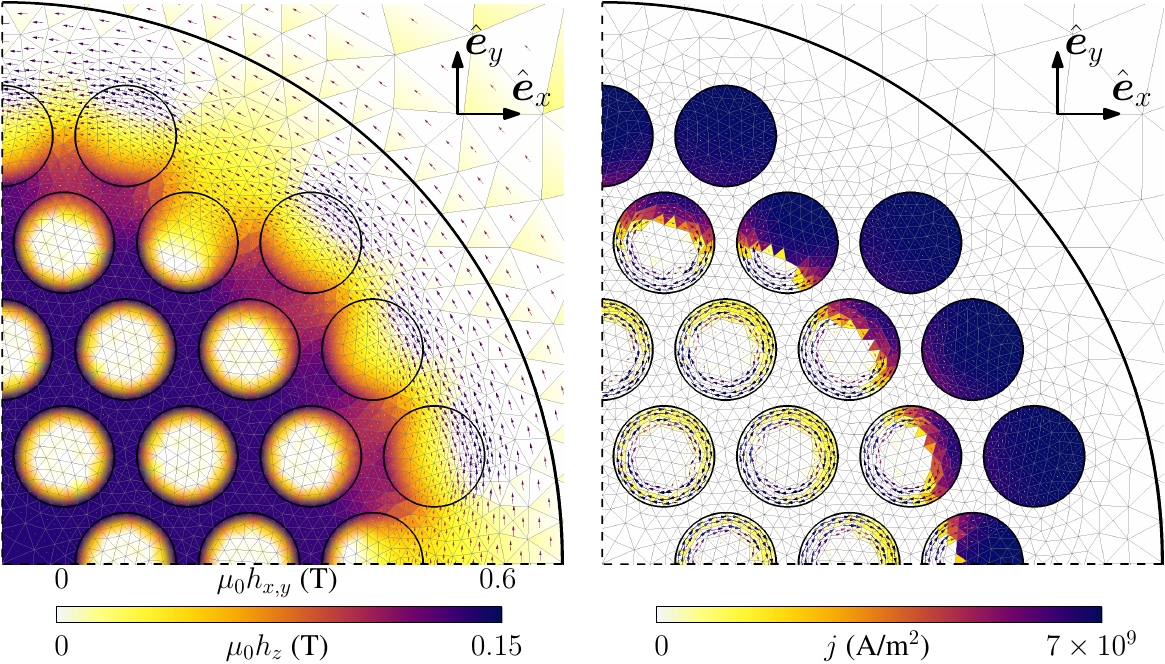}
\caption{Magnetic field (left) and current density (right) for the 54-filament wire at time $t=0.1$ s. Only one quarter of the geometry is shown. Arrows represent the in-plane components and the elements are colored as a function of the value of the $z$-component of the vectors. Note that two different scales are used for the magnetic field for clarity.}
\label{54_fil_solution}
\end{center}
\end{figure}

Fig.~\ref{fil_54_current} shows the time evolution of the current in the filaments depending on their position. As expected, the current density progressively penetrates into inner layers of the wire. Due to the twist, the current flowing in the outer filaments generates a non-zero $h_z$ component inside the wire. Circulating in-plane currents therefore appear in the inner layers to shield this axial magnetic field, as illustrated in Fig.~\ref{54_fil_solution}. 

Note that during the first transport current increase (for $t<T/4$), no filament carries a negative current. This is in contradiction to what was obtained in~\cite{tominaka2005calculations} with an alternative method on a similar problem, where a critical state model is considered and the current density is assumed localized along a line within every filament, which therefore leads to neglecting the spatial extension of the filaments.

%

\FloatBarrier

\section{Extension to Non-Helicoidally-Invariant Boundary Conditions \textemdash\   General BC}\label{sec_helicoidal_non_symmetric}

\begin{figure}[h!]
\begin{center}
\includegraphics[width=\linewidth]{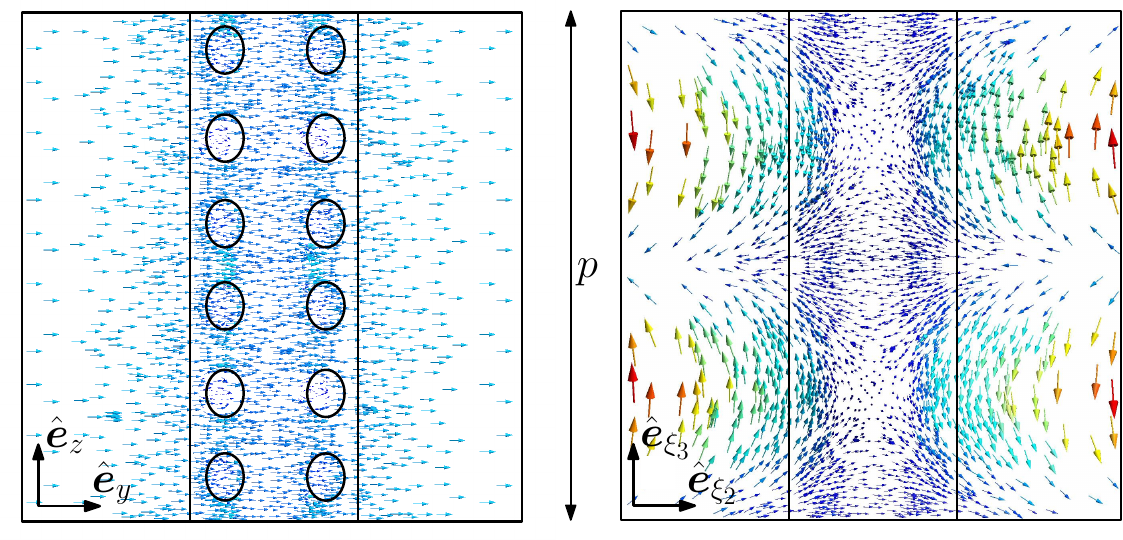}
\caption{Uniform magnetic field along $\ey$ for the six-filament geometry represented in Fig.~\ref{filamentGeometry_full}. (Left) In the physical space on the plane $x=0$. (Right) In the helicoidal coordinate system represented as an orthogonal system on the plane $\xi_1=0$.}
\label{twistedConstantField}
\end{center}
\end{figure}

When BC are not HS, the dimension of the problem cannot be directly reduced from 3D to 2D on basis of the geometrical symmetry only. This is the case when a wire is subjected to a uniform transverse magnetic field. For an applied magnetic field $\h_x = (0\ 1\ 0)\transpose$, we have
\begin{align}\label{eq_transverseField}
\h_{\vec \xi} = \mat J^{\text{T}}\ \h_{\vec x} = \begin{pmatrix}
\sin \alpha \xi_3\\
\cos \alpha \xi_3\\
\alpha \xi_1 \cos \alpha \xi_3 - \alpha \xi_2 \sin \alpha \xi_3
\end{pmatrix},
\end{align}
which is not $\xi_3$-independent, see Fig.~\ref{twistedConstantField}. As a consequence, the solution of the magnetodynamic problem will not be $\xi_3$-independent either. The periodic structure of the problem can however be exploited by expressing the solution as a series of periodic functions with respect to $\xi_3$.


In this section, we present this approach and show that it generalizes the method described in Section~\ref{sec_helicoidal_symmetric2D}. In particular, we show that it also leads to a 2D model in helicoidal coordinates, which has the potential of considerably reducing the computational cost compared to a 3D model.

\subsection{Fourier decomposition of the magnetic field}

Given the $p$-periodicity with respect to $\xi_3$, by separation of variables, we can expand the magnetic field $\h=\h(\xi_1,\xi_2,\xi_3)$ in the following series:
\begin{align}\label{eq_h_helicoidal_fullExpansion_fullField}
\h(\xi_1,\xi_2,\xi_3) &= \sum_{k=-\infty}^{\infty} \h_k(\xi_1,\xi_2) f_k(\xi_3),
\end{align}
with the \textit{modes} $f_k = f_k(\xi_3)$ that are functions of $\xi_3$ only and that are defined as
\begin{equation}\label{eqn_modeDefinition}
f_{k}(\xi_3) = \left\{\begin{aligned}
&\sqrt{2}\cos(\alpha k \xi_3), \  &k<0,\\
&1, \ &k=0,\\
&\sqrt{2}\sin(\alpha k \xi_3), \  &k>0,
\end{aligned} \right.
\end{equation}
and with the \textit{spatial Fourier coefficients} $\h_k = \h_k(\xi_1,\xi_2)$ that are three-component vector functions of $\xi_1$ and $\xi_2$.

The modes $f_k$ are mutually orthogonal and have a unit norm, denoted as $\|f_k\|=1$, in the sense of the following inner product:
\begin{align}\label{eq_innerProduct}
\langle f_{k_1} , f_{k_2} \rangle = \frac{1}{p} \int_0^p f_{k_1} f_{k_2}\ d\xi_3= \delta_{k_1 k_2},\quad \forall k_1,k_2\in \mathbb{Z}.
\end{align}
They also satisfy the following property:
\begin{align}\label{modesProperty}
\derd{f_k}{\xi_3} = \alpha k f_{-k}, \quad \forall k\in \mathbb{Z}.
\end{align}

Introducing a decomposition of the magnetic field into its in-plane and out-of-plane components as in the HS case of Section~\ref{sec_helicoidal_symmetric2D} for each Fourier coefficient $\h_k$, we can rewrite Eq.~\eqref{eq_h_helicoidal_fullExpansion_fullField} as
\begin{align}\label{eq_h_helicoidal_fullExpansion}
\h &= \sum_{k=-\infty}^{\infty} \paren{\hpak{k}(\xi_1,\xi_2) + \hpek{k}(\xi_1,\xi_2)} f_k(\xi_3),
\end{align}
with the $\hpak{k}$ containing the $\xi_1$ and $\xi_2$-components of $\h_k$, and the $\hpek{k}$ containing its $\xi_3$-component. Equation~\eqref{eq_h_helicoidal_fullExpansion} actually generalizes the decomposition in Eq.~\eqref{eq_h_decomposition} for the case of HS BC. Indeed, in the case of HS BC, the only mode that is involved is $f_0(\xi_3) = 1$, with coefficients $\hpak{0}=\hpa$ and $\hpek{0}=\hpe$, and the coefficients of the other modes, $\hpak{k}$ and $\hpek{k}$, $\forall k\in \mathbb{Z}_0$, are all equal to zero.

\subsection{Space discretization with curl-free functions in $\Occ$}

The curl of decomposition \eqref{eq_h_helicoidal_fullExpansion} reads
\begin{align}\label{eq_curlDecomposition}
\curl \h = \sum_{k=-\infty}^{\infty}\Big( f_k\ \curl \hpak{k} + \derd{f_k}{\xi_3} \ezh \times \hpak{k}& \notag\\
+ f_k\ \curl\hpek{k}&\Big),
\end{align}
where $\ezh$ is the unit vector in the $\xi_3$-direction.

The only term in Eq.~\eqref{eq_curlDecomposition} contributing to the $\xi_3$-component of the curl involves the curl of $\hpak{k}$. We can therefore keep the same discrete function space for the $\hpak{k}$ as for $\hpa$ in Section~\ref{sec_helicoidal_symmetric2D}, however without the $\sum_i I_i \vec c_i$ term of Eq.~\eqref{eq_h_parallel_decomposition} for $k\neq 0$, as transport currents only contribute to the fundamental mode with $f_0(\xi_3) = 1$. 

As in the $\xi_3$-independent case, we express the out-of-plane magnetic field $\hpek{k}$ as a sum of perpendicular edge functions. But now, the $\hpak{k}$ functions also contribute to the $\xi_1$ and $\xi_2$-components of the curl of $\h$ for $k\neq 0$ in Eq.~\eqref{eq_curlDecomposition} via the cross product term. Therefore, the curl-free condition in $\Occ$ is no longer met with a uniform out-of-plane magnetic field in $\Occ$, for $k\neq 0$. Instead, as is shown below, the curl-free condition induces a coupling between the in-plane and out-of-plane magnetic field contributions in $\Occ$. For simplicity, as was done before, we assume that there is only one connected non-conducting region $\Occ$.

Using curl-free in-plane functions $\hpak{k}$ in $\Occ$ and Eq.~\eqref{eq_curlDecomposition}, the curl-free condition on $\h$ in $\Occ$ reads
\begin{align}
\sum_{k=-\infty}^{\infty}\Big( \derd{f_k}{\xi_3} \ezh \times \hpak{k}\ +\ f_k\ \curl\hpek{k}\Big) = \vec 0.
\end{align}
Using the mode property Eq.~\eqref{modesProperty}, this yields
\begin{align}
\sum_{k=-\infty}^{\infty} \Big(&\curl\hpek{k} -\alpha k \ezh \times \hpak{-k}\Big)f_{k} = \vec 0,
\end{align}
which results in the following condition, $\forall k\in \mathbb{Z}$:
\begin{align}\label{eq_curlFreeCondition_component}
\curl\hpek{k} -\alpha k \ezh \times \hpak{-k} = \vec 0.
\end{align}
For $k = 0$, we retrieve the same condition as in the helicoidally symmetric problem, that is, $\hpek{0}$ must be uniform in $\Occ$, with a value given by Eq.~\eqref{eq_hperp_condition}. For $k\neq 0$, the condition can be enforced via the independent degrees of freedom of the in-plane and out-of-plane magnetic field contributions. Indeed, in $\Occ$, we have the expansions
\begin{align}
\hpek{k} &= \sum_{n\in \nodes(\Occ)} \phipek{k,n}\ w_n \ezh,\\
\hpak{-k} &= \sum_{n\in \nodes(\Occ)} \phipak{-k,n} \ \grad w_n,
\end{align}
where $w_n \ezh = \vec w_n$ is the perpendicular edge function of node $n$, with $w_n$ the usual node function. In terms of the individual degrees of freedom, Eq.~\eqref{eq_curlFreeCondition_component} reads
\begin{align}
& \sum_{n\in \nodes(\Occ)} \paren{\phipek{k,n} + \alpha k \ \phipak{-k,n}}\begin{pmatrix}
\partial_{\xi_2} w_n\\
-\partial_{\xi_1} w_n\\
0
\end{pmatrix} =\vec 0.
\end{align}
This equation is valid over the whole domain $\Occ$ if and only if the first parenthesis is constant. This is the case if, for $k\neq 0$,
\begin{align}\label{eq_conditiononcoefficients}
\phipek{k,n} + \alpha k \ \phipak{-k,n} = 0,\ \forall n\in \nodes(\Occ).
\end{align}
That is, to ensure a curl-free magnetic field in $\Occ$, the degrees of freedom of the mode $\hpek{k}$ must be linked directly to those of the mode $\hpak{-k}$ in $\Occ$ (or vice-versa). 

This link between the degrees of freedom strongly ensures that $\curl \h=\vec 0$ in $\Occ$ and allows for a significant reduction of the number of unknowns, hence a reduction of the computational cost of the resolutions.

\subsection{Boundary conditions for a transverse magnetic field}

The transverse magnetic field defined in Eq.~\eqref{eq_transverseField} applied as a BC on $\Gout$ only involves the modes $f_{-1}(\xi_3)$ and $f_1(\xi_3)$. We have, in helicoidal components:
\begin{align}
&(\hpak{-1})_{\vec{\xi}} = \frac{\sqrt{2}}{2}\begin{pmatrix}
0 & 1 & 0
\end{pmatrix}\transpose,\label{eq_bc_transverse_1}\\
&(\hpek{-1})_{\vec{\xi}} = \frac{\sqrt{2}}{2}\begin{pmatrix}
0 & 0 & \alpha \xi_1
\end{pmatrix}\transpose,\\
&(\hpak{+1})_{\vec{\xi}} = \frac{\sqrt{2}}{2} \begin{pmatrix}
1 & 0 & 0
\end{pmatrix}\transpose,\\ 
&(\hpek{+1})_{\vec{\xi}} = \frac{\sqrt{2}}{2}\begin{pmatrix}
0 & 0 & -\alpha \xi_2
\end{pmatrix}\transpose.
\label{eq_bc_transverse_2}
\end{align}
We can verify that they satisfy Eq.~\eqref{eq_curlFreeCondition_component}.


\subsection{Derivation of the \hpf with linear materials}

With linear materials, orthogonality allows solving modes with different values of $|k|$ (i.e., including $-k$ and $k$) independently. For each value of $|k|$, the integration along $\xi_3$ gives an independent set of equations, written in terms of the unknown Fourier coefficients $\hpak{-k}$, $\hpek{-k}$, $\hpak{k}$, and $\hpek{k}$. These coefficients are functions of $\xi_1$ and $\xi_2$ only, and hence the problem is 2D. The formulation is derived in the Appendix.

For nonlinear materials, the modes are no longer decoupled. We provide observations and comments on how to handle this situation in Section~\ref{sec_commentNonlinear}.

\section{Verification and Application \textemdash\ General BC}\label{sec_filaments_verifquasi3Dmodel}

In this section, we first verify the implementation of the generalized 2D-$\xi$ method by comparing its results with those of a 3D reference model, for linear materials. We then apply the method on a 54-filament wire and discuss the different contributions to the total AC loss, still with linear materials. Finally, we comment on the application of the method in the case of nonlinear materials.

\subsection{Verification with linear materials}

The validity of the approach with linear materials is verified by comparing the results of the \tdpk\ model with those obtained with a classical 3D model. We consider the same geometry as in Section~\ref{sec_filaments_verif3Dmodel}, but with a constant resistivity in the filaments, and with a uniform transverse magnetic field instead of an imposed transport current.

The filaments have a constant resistivity $\rho_\text{SC} = 3.3 \times 10^{-14}$ $\Omega$m (dummy value chosen for verification), and the matrix has a constant resistivity $\rho_\text{Cu} = 1.81 \times 10^{-10}$ $\Omega$m. The system is subjected to a transverse magnetic field along $y$, increasing from $0$ T to $0.1$ T with a constant ramp-up rate of $18$~T/s.

\begin{figure}[h!]
\begin{center}
\includegraphics[width=\linewidth]{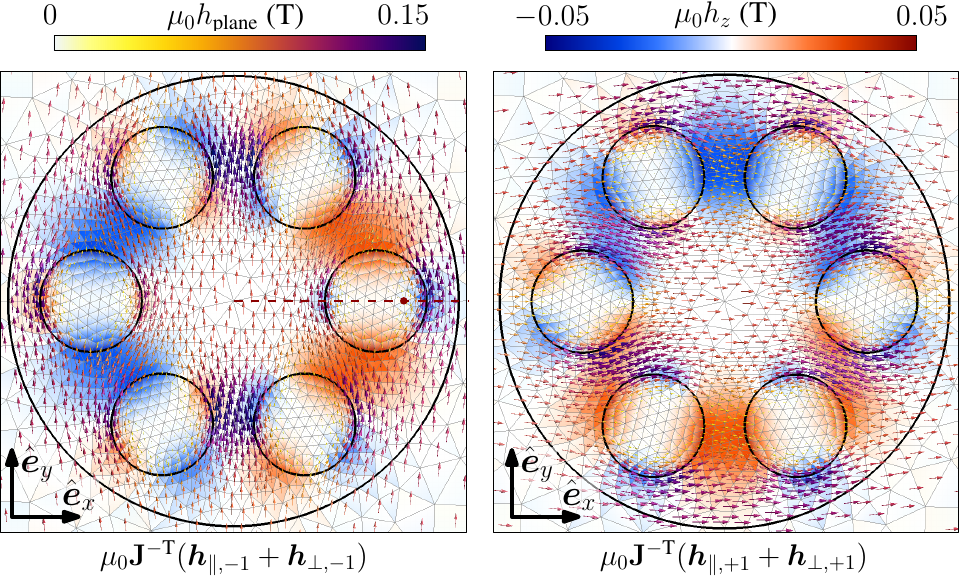}
\caption{Solution of the \tdpk\ model with linear materials, on the $z=0$ plane, for a transverse field of $0.1$~T. The arrows represent $\mu_0 \h$, and the triangular elements are colored as a function of the value of $\mu_0 h_z$, using the color map on the top. The dashed red line is where the field is taken for Fig.~\ref{transverse_b_line_verif}, and the red dot along that line represents the intersection with the plane $z=0$ of the helicoidal fiber along which the field is taken for Fig.~\ref{filament_verif_helicoidal_3D_transverse}. (Left) Mode with $f_{-1}(\xi_3)$ (see Eq.~\eqref{eq_linkOcc_transverse}). (Right) Mode with $f_{+1}(\xi_3)$ (see Eq.~\eqref{eq_linkOcc_transverse}).}
\label{filament_transverse_bcos_bsin}
\end{center}
\end{figure}

Boundary conditions for the \tdpk\ model are imposed on $\Gout$ so as to satisfy Eqn.~\eqref{eq_bc_transverse_1} to \eqref{eq_bc_transverse_2}. Only modes $f_{-1}(\xi_3) =\sqrt{2} \cos \alpha \xi_3$ and $f_{+1}(\xi_3) = \sqrt{2} \sin \alpha \xi_3$ are therefore excited so that the full magnetic field reads
\begin{align}\label{eq_linkOcc_transverse}
\h = \paren{\hpak{-1}(\xi_1,\xi_2) + \hpek{-1}(\xi_1,\xi_2)} f_{-1}(\xi_3)&\notag\\
+ \paren{\hpak{+1}(\xi_1,\xi_2) + \hpek{+1}(\xi_1,\xi_2)} f_{+1}(\xi_3)&.
\end{align}
The result of the linear \tdpk\ model is illustrated in Fig.~\ref{filament_transverse_bcos_bsin}.

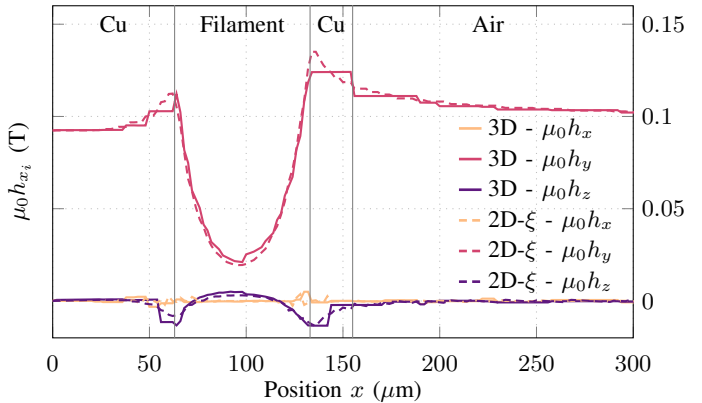
\begin{figure}[h!]
        \centering
             \tikzsetnextfilename{transverse_b_line_verif}
\begin{tikzpicture}[trim axis left, trim axis right][font=\small]
 	\begin{axis}[
    width=1.05\linewidth,
    height=6cm,
    grid = both,
    grid style = dotted,
    ymin=-0.02, 
    ymax=0.16,
    xmin=0, 
    xmax=300,
     yticklabel style={xshift=-0.4em},
    yticklabels={$0$, $0$, $0.05$, $0.1$, $0.15$},
    xticklabel style={yshift=-0.4em},
	xlabel={Position $x$ ($\mu$m)},
    ylabel={$\mu_0 h_{x_i}$ (T)},
    ylabel style={yshift=-2.5em},
    yticklabel style={xshift=0.4em},
    yticklabel pos=right,
    xlabel style={yshift=0.2em},
    legend style={at={(0.99, 0.40)}, cells={anchor=west}, anchor= east, draw=none, fill opacity=0, text opacity = 1, legend image code/.code={\draw[##1,line width=1pt] plot coordinates {(0cm,0cm) (0.3cm,0cm)};}}
    ]
    \addplot[mag_6, thick] 
    table[x=r,y=bx_3D]{data/transverse_b_line_verif.txt};
    \addplot[mag_4, thick] 
    table[x=r,y=by_3D]{data/transverse_b_line_verif.txt};
    \addplot[mag_2, thick]
    table[x=r,y=bz_3D]{data/transverse_b_line_verif.txt};
        \addplot[mag_6, thick, densely dashed] 
    table[x=r,y=bx_2D]{data/transverse_b_line_verif.txt};
    \addplot[mag_4, thick, densely dashed] 
    table[x=r,y=by_2D]{data/transverse_b_line_verif.txt};
        \addplot[mag_2, thick, densely dashed] 
    table[x=r,y=bz_2D]{data/transverse_b_line_verif.txt};
     \addplot[gray] 
    coordinates {(63,-1) (63, 1)}; 
         \addplot[gray] 
    coordinates {(133,-1) (133, 1)};
    \addplot[gray] 
     coordinates {(155,-1) (155, 1)}; 
     \node[] at (axis cs: 31.5,0.15) {Cu};
     \node[] at (axis cs: 98,0.15) {Filament};
    	 \node[] at (axis cs: 144,0.15) {Cu};
     \node[] at (axis cs: 225,0.15) {Air};
    \legend{3D - $\mu_0 h_x$, 3D - $\mu_0 h_y$, 3D - $\mu_0 h_z$, \tdpk\ - $\mu_0 h_x$, \tdpk\ - $\mu_0 h_y$, \tdpk\ - $\mu_0 h_z$}
    \end{axis}
\end{tikzpicture}%
\vspace{-0.2cm}
\caption{Magnetic field components along the dashed red line represented in Fig.~\ref{filament_transverse_bcos_bsin}, at $z=0$, for the 3D and \tdpk\ models with a fine mesh resolution for $\mu_0 h_y=0.1$~T.}
        \label{transverse_b_line_verif}
\end{figure}

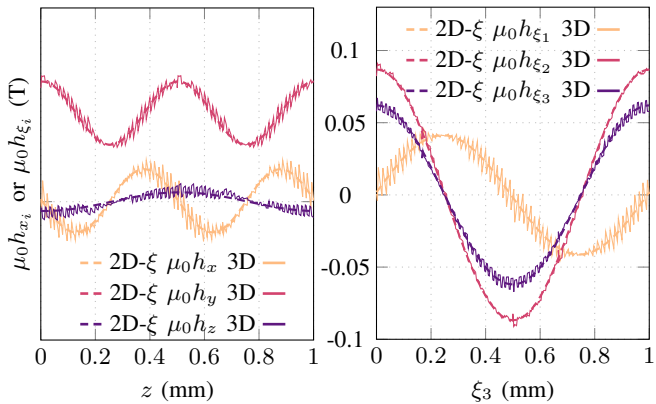
\begin{figure}[h!]
        \centering
        \begin{subfigure}[b]{0.49\linewidth}
        \centering
             \tikzsetnextfilename{filament_verif_helicoidal_3D_transverse_x}
	\begin{tikzpicture}[trim axis left, trim axis right][font=\small]
	\begin{axis}[
    width=1.2\textwidth,
    height=6cm,
    grid = both,
    grid style = dotted,
    xmin=0, 
    xmax=1,
    ymin=-0.1, 
    ymax=0.14,
    yticklabel=\empty,
	xtick scale label code/.code={},
    ytick={-0.5, 0, 0.5, 1, 1.5, 2},
    yticklabel style={anchor=center},
	ytick scale label code/.code={},
	xlabel={$z$ (mm)},
    ylabel={$\mu_0 h_{x_i}$ or $\mu_0 h_{\xi_i}$ (T)},
    ylabel style={yshift=-3em},
    xlabel style={yshift=0.3em},
    legend columns=2,
    legend style={at={(0.54, 0.0)}, anchor=south, style={column sep=0cm}, inner sep=0pt, draw=none, fill opacity=0, text opacity = 1, legend image code/.code={\draw[##1,line width=1pt] plot coordinates {(0cm,0cm) (0.3cm,0cm)};}}
    ]
    \addplot[mag_6, thick, densely dashed] 
    table[x=z,y=bx_2D]{data/filament_verif_helicoidal_3D_transverse.txt};
        \addplot[mag_6, thin] 
    table[x=z,y=bx]{data/filament_verif_helicoidal_3D_transverse.txt};
    \addplot[mag_4, thick, densely dashed] 
    table[x=z,y=by_2D]{data/filament_verif_helicoidal_3D_transverse.txt};
        \addplot[mag_4, thin] 
    table[x=z,y=by]{data/filament_verif_helicoidal_3D_transverse.txt};
        \addplot[mag_2, thick, densely dashed] 
    table[x=z,y=bz_2D]{data/filament_verif_helicoidal_3D_transverse.txt};
            \addplot[mag_2, thin] 
    table[x=z,y=bz]{data/filament_verif_helicoidal_3D_transverse.txt};
    \legend{\tdpk\ $\mu_0 h_{x}$ 3D, ~,\tdpk\ $\mu_0 h_{y}$ 3D, ~,\tdpk\ $\mu_0 h_{z}$ 3D, ~}
    \end{axis}
	\end{tikzpicture}%
        \label{filament_verif_helicoidal_3D_transverse_x}
	        \end{subfigure}
        \begin{subfigure}[b]{0.49\linewidth}  
\centering
             \tikzsetnextfilename{filament_verif_helicoidal_3D_transverse_xi}
	\begin{tikzpicture}[trim axis left, trim axis right][font=\small]
		\begin{axis}[
    width=1.2\textwidth,
    height=6cm,
    grid = both,
    grid style = dotted,
    xmin=0, 
    xmax=1,
    ymin=-0.1, 
    ymax=0.13,
	xtick scale label code/.code={},
    ytick={-0.1, -0.05, 0, 0.05, 0.1},
    yticklabels={-0.1, -0.05, 0, 0.05, 0.1},
    yticklabel style={anchor=center},
	ytick scale label code/.code={},
	xlabel={$\xi_3$ (mm)},
    ylabel style={yshift=-0.5em},
    xlabel style={yshift=0.3em},
    yticklabel style={anchor=center},
    yticklabel style={xshift=-1.3em},
    legend columns=2,
    legend style={at={(0.52, 0.99)}, anchor=north, style={column sep=0cm}, inner sep=1.2pt, draw=none, fill opacity=0, text opacity = 1, legend image code/.code={\draw[##1,line width=1pt] plot coordinates {(0cm,0cm) (0.3cm,0cm)};}}
    ]
    \addplot[mag_6, thick, densely dashed] 
    table[x=z,y=bxi1_2D]{data/filament_verif_helicoidal_3D_transverse.txt};
        \addplot[mag_6, thin] 
    table[x=z,y=bxi1]{data/filament_verif_helicoidal_3D_transverse.txt};
    \addplot[mag_4, thick, densely dashed] 
    table[x=z,y=bxi2_2D]{data/filament_verif_helicoidal_3D_transverse.txt};
        \addplot[mag_4, thin] 
    table[x=z,y=bxi2]{data/filament_verif_helicoidal_3D_transverse.txt};
        \addplot[mag_2, thick, densely dashed] 
    table[x=z,y=bxi3_2D]{data/filament_verif_helicoidal_3D_transverse.txt};
            \addplot[mag_2, thin] 
    table[x=z,y=bxi3]{data/filament_verif_helicoidal_3D_transverse.txt};
    \legend{\tdpk\ $\mu_0 h_{\xi_1}$ 3D, ~, \tdpk\ $\mu_0 h_{\xi_2}$ 3D, ~, \tdpk\ $\mu_0 h_{\xi_3}$ 3D, ~}
    \end{axis}
	\end{tikzpicture}%
        \label{filament_verif_helicoidal_3D_transverse_xi}
            \end{subfigure}
        \caption{Magnetic field along the helicoidal fiber of pitch length $p$, passing at point $\vec x = \big(r, 0, 0\big)$, with $r=R_\ell + 0.8 R_\text{f}$ (represented by the red dot in Fig.~\ref{filament_transverse_bcos_bsin}), from $z=0$ to $z=p$, for the 3D and \tdpk\ models for $\mu_0 h_y=0.1$~T. (Left) Cartesian components of the vectors. (Right) Helicoidal components of the vectors.}
        \label{filament_verif_helicoidal_3D_transverse}
\end{figure}

Comparisons with the solution of the 3D problem are given in Figs.~\ref{transverse_b_line_verif} and \ref{filament_verif_helicoidal_3D_transverse}, along a characteristic line in the $z=\xi_3=0$ plane and along a helicoidal fiber of pitch length $p$, passing at point $\vec x = \big(r, 0, 0\big)$, with $r=R_\ell + 0.8 R_\text{f}$, from $z=0$ to $z=p$. Both models agree with each other.

As in the HS-BC case, exploiting the geometrical symmetry allows for a strong reduction of the computational work. It should however be mentioned that the \tdpk\ model with transverse field BC involves double number of degrees of freedom compared to the same model with HS-BC, as two modes are needed to represent the transverse field ($-k$ and $k$, compared to $k=0$ only). 

The \tdpk\ model still leads to a considerable reduction of DOFs compared to the 3D model. Indeed, taking values of the fine mesh resolution from Table.~\ref{filament_efficiency}, the 3D model involves 145k DOFs whereas the \tdpk\ model with two modes only involves 12k DOFs.

\subsection{Application on a 54-filament wire with linear materials}

We now consider the 54-filament geometry defined in Fig.~\ref{54fil_appCurrent} but with a linear material in the filaments. We fix the resistivity in the filaments to $\rho_\text{SC} = 1.81\times 10^{-15}$ $\Omega$m and in the matrix to $\rho_\text{Cu} = 1.81\times 10^{-10}$ $\Omega$m.

Choosing such a low resistivity in filaments leads to an approximated model for superconducting wires at low magnetic fields (below filament saturation). We will see that this linear model reproduces the coupling current dynamics observed in the copper matrix of superconducting wires. The validity of the linear model is however limited to this. It does not describe superconducting hysteresis effects in filaments, and hence does not allow for superconducting loss calculation. Instead, in the following, the computed losses in the filament region will be those of a normal resistive material.

We compute the AC loss induced by a time-varying transverse magnetic field $\mu_0 \h = b_\text{max} \sin(\omega t)\ey$, with $b_\text{max}=0.1$~T, as a function of the frequency $f = \omega / 2\pi$. As in the previous section, BC are such that only the modes $f_{-1}(\xi_3)$ and $f_{+1}(\xi_3)$ are excited (see Eqn.~\eqref{eq_bc_transverse_1} to \eqref{eq_bc_transverse_2}).

Moreover, because the materials are linear and the excitation is harmonic, the problem can be solved in the frequency domain. To this end, we write the problem in terms of the auxiliary complex quantity $\hat \h(\vec \xi)$, the phasor of the magnetic field. The phasor is related to the physical magnetic field by $\h(\vec \xi, t) = \Re\big(\hat \h(\vec \xi)e^{i\omega t}\big)$, with $i = \sqrt{-1}$, and we replace all time derivatives in the formulation by a multiplication by $i\omega$.

The time-average instantaneous loss density, in W/m$^3$, reads, in terms of Cartesian and helicoidal components of the phasor $\hat \j$ for the current density:
\begin{align}\label{eq_avgLossDensity}
\frac{1}{2}\,\hat \j^\star_{\vec x}\, \paren{\rho\, \hat \j_{\vec x}} = \frac{1}{2}\, \hat \j^\star_{\vec \xi}\, \rhotilde\, \hat \j_{\vec \xi},
\end{align}
where $\hat\j^\star$ denotes the tranposed complex conjugate of $\hat \j$. Note that both sides of Eq.~\eqref{eq_avgLossDensity} are real since $\rho$ is a scalar and $\rhotilde$ is a hermitian tensor. The \textit{total loss} per unit length is obtained by integrating Eq.~\eqref{eq_avgLossDensity} over the whole wire cross section.

We decompose this total loss into separate contributions, which allows for an easier interpretation of the results. The \textit{filament loss} is the integral of Eq.~\eqref{eq_avgLossDensity} on the filament region only. The \textit{coupling loss} is the integral of Eq.~\eqref{eq_avgLossDensity} on the matrix region only, taking only the in-plane components of $\hat \j_{\vec x}$ into account (the $x$ and $y$ Cartesian components). Finally, the \textit{eddy current loss} is the same but with only the out-of-plane component $\hat j_{z}$.

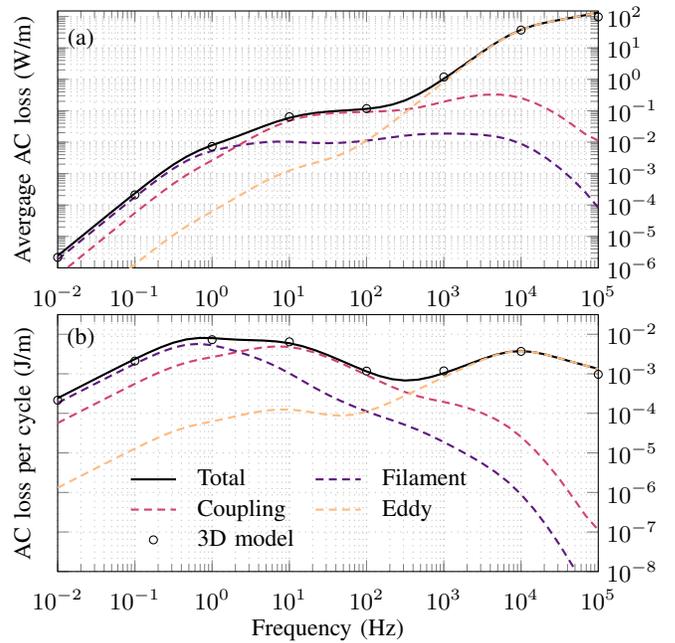
\begin{figure}[h!]
        \centering
\begin{subfigure}[b]{0.99\linewidth}  
\centering
\tikzsetnextfilename{fil54_transverse_10mm_total}
\begin{tikzpicture}[trim axis left, trim axis right][font=\small]
\pgfplotsset{set layers}
 	\begin{loglogaxis}[
	tick label style={/pgf/number format/fixed},
    width=\linewidth,
    height=5cm,
    grid = both,
    grid style = dotted,
    ymin=1e-6, 
    ymax=1.5e2,
    xmin=0.01, 
    xmax=100000,
    ytick={1e-6, 1e-5, 1e-4, 1e-3, 1e-2, 1e-1, 1e0, 1e1, 1e2},
    ylabel={Avergage AC loss (W/m)},
    ylabel style={yshift=-2.5em},
    ylabel style={xshift=0em},
    xlabel style={yshift=0.4em},
    xticklabel style={yshift=-0.4em},
    yticklabel style={xshift=0em},
    yticklabel pos=right,
    legend style={at={(0.5, 0.05)}, cells={anchor=west}, anchor=south, draw=none,fill opacity=0, text opacity = 1}
    ]
    \addplot[black, thick] 
    table[x=f,y=total]{data/I_54fil_transverse_10mm_veryfine.txt};
        \addplot[mag_2, densely dashed, thick] 
    table[x=f,y=fil]{data/I_54fil_transverse_10mm_veryfine.txt};
        \addplot[mag_4, densely dashed, thick] 
    table[x=f,y=coupling]{data/I_54fil_transverse_10mm_veryfine.txt};
        \addplot[mag_6, densely dashed, thick] 
    table[x=f,y=eddy]{data/I_54fil_transverse_10mm_veryfine.txt};
        \addplot[only marks, color=black, mark=o,mark size=1.5pt] 
    table[x=f,y=total]{data/I_54fil_transverse_10mm_3D.txt};
\node[] at (axis cs: 0.02, 20) {(a)};
    \end{loglogaxis}
\end{tikzpicture}%
\end{subfigure}
        \hfill
\begin{subfigure}[b]{0.99\linewidth}  
\centering
\tikzsetnextfilename{fil54_transverse_10mm_percycle}
\begin{tikzpicture}[trim axis left, trim axis right][font=\small]
\pgfplotsset{set layers}
 	\begin{loglogaxis}[
	tick label style={/pgf/number format/fixed},
    width=\linewidth,
    height=5cm,
    grid = both,
    grid style = dotted,
    ymin=1e-8, 
    xmin=0.01, 
    xmax=100000,
    ytick={1e-7, 1e-8, 1e-6, 1e-5, 1e-4, 1e-3, 1e-2},
	xlabel={Frequency (Hz)},
    ylabel={AC loss per cycle (J/m)},
    ylabel style={yshift=-2.5em},
    xlabel style={yshift=0em},
    xticklabel style={yshift=-0.4em},
    yticklabel style={xshift=0em},
    yticklabel pos=right,
    legend columns=2,
    legend style={at={(0.45, 0.05)}, cells={anchor=west}, style={column sep=0.2cm}, inner sep=1.2pt, anchor=south, draw=none,fill opacity=0, text opacity = 1}
    ]
    \addplot[black, thick] 
    table[x=f,y=total_f]{data/I_54fil_transverse_10mm_veryfine.txt};
        \addplot[mag_2, densely dashed, thick] 
    table[x=f,y=fil_f]{data/I_54fil_transverse_10mm_veryfine.txt};
        \addplot[mag_4, densely dashed, thick] 
    table[x=f,y=coupling_f]{data/I_54fil_transverse_10mm_veryfine.txt};
        \addplot[mag_6, densely dashed, thick] 
    table[x=f,y=eddy_f]{data/I_54fil_transverse_10mm_veryfine.txt};
            \addplot[only marks, color=black, mark=o,mark size=1.5pt] 
    table[x=f,y=total_f]{data/I_54fil_transverse_10mm_3D.txt};
    \legend{Total, Filament, Coupling, Eddy, 3D model}
        \node[] at (axis cs: 0.02, 0.008) {(b)};
    \end{loglogaxis}
\end{tikzpicture}%
\end{subfigure}
\vspace{-0.2cm}
\caption{AC loss as a function of the frequency of an external transverse magnetic field for linear materials and $p=10$~mm. The total loss as well as separate contributions (filament, coupling, eddy) are shown. The legend is valid for both subfigures. The markers denote the total loss obtained by a 3D model, for verification.}
        \label{fil54_transverse_10mm}
\end{figure}

\begin{figure}[h!]
\begin{center}
\includegraphics[width=\linewidth]{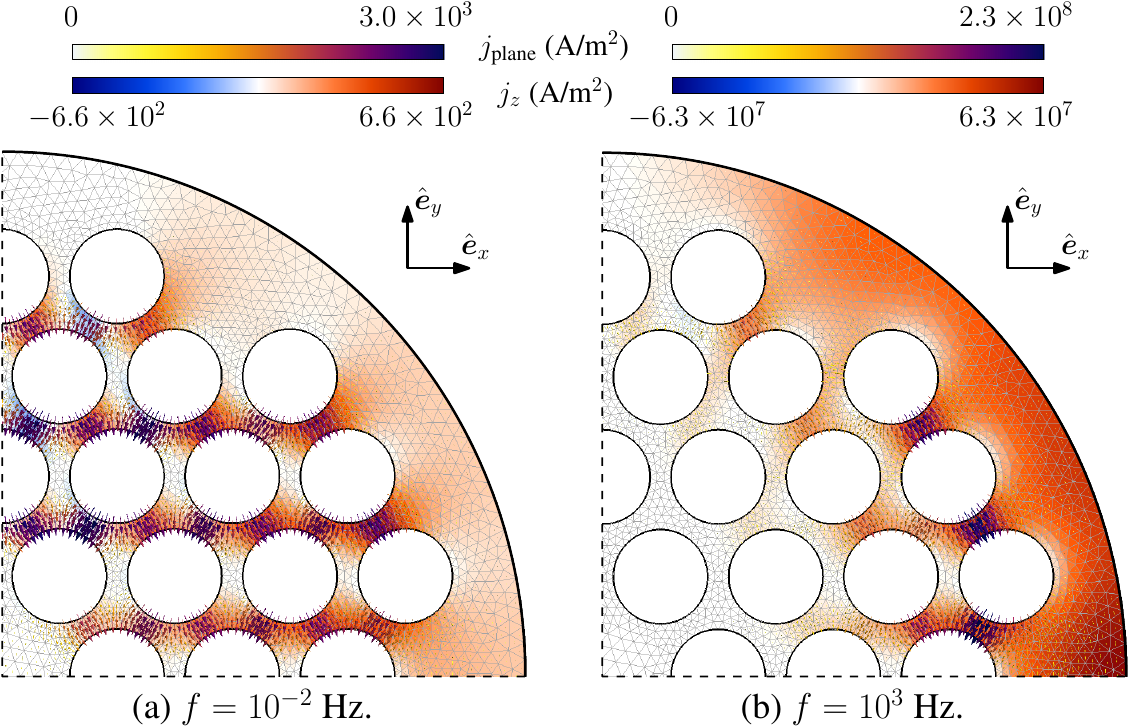}
\caption{Real part of the current density distribution in the matrix in transverse magnetic field in harmonic regime. Arrows represent the in-plane $x$-$y$-components, and elements are colored as a function of the out-of-plane $z$-component.}
\label{54_fil_transverse_sol}
\end{center}
\end{figure}

We present the results for frequencies ranging from $10^{-2}$~Hz to $10^{5}$ Hz in Fig.~\ref{fil54_transverse_10mm} for a pitch length $p=10$~mm. Values from a 3D reference model are also given for comparison, the agreement with the \tdpk\ model is very good. The current distribution at two distinct frequencies is shown in Fig.~\ref{54_fil_transverse_sol}.

\begin{figure}[h!]
\begin{center}
\includegraphics[width=\linewidth]{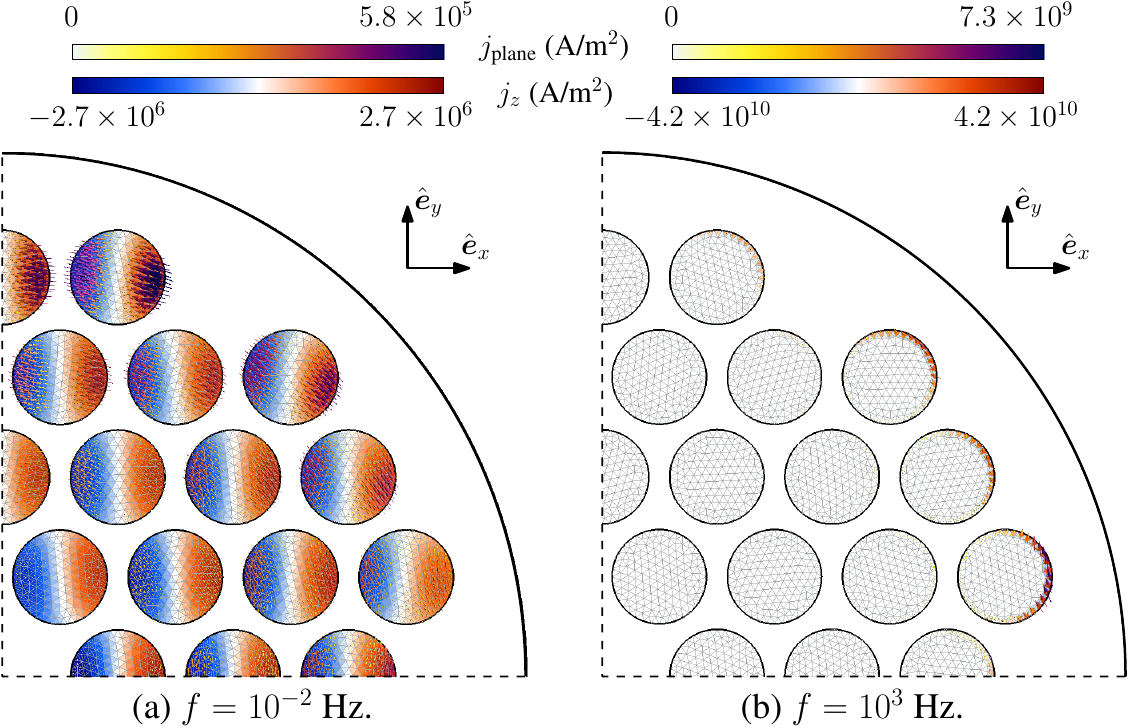}
\caption{Real part of the current distribution in the filaments in transverse magnetic in harmonic regime. Arrows represent the in-plane $x$-$y$-components, and elements are colored as a function of the out-of-plane $z$-component.}
\label{54_fil_transverse_sol_full_fil}
\end{center}
\end{figure}

Figure~\ref{fil54_transverse_10mm} shows that the dominant loss contribution depends on the frequency. This can be interpreted as follows.


Neglecting the effect of the twist, we expect the peak value of the filament loss to arise when the diffusion skin depth $\delta_\text{SC} = \sqrt{2\rho_\text{SC} /\omega \mu}$ is comparable with the radius of the wire $R_\text{f}$. We have $\delta_\text{SC} / R_\text{f} = 1$ for the frequency $f = 0.22$~Hz, which is not too far from the peak in Fig.~\ref{fil54_transverse_10mm}(b).

A change of regime for the eddy current loss per cycle should arise when the skin effect starts to play a role in the matrix. This is expected to happen when the diffusion skin depth $\delta_\text{Cu} = \sqrt{2\rho_\text{Cu} /\omega \mu}$ is comparable with the thickness of the outer sheath of the matrix $D_\text{os}\approx 80$ $\mu$m. Here, we have $\delta_\text{Cu} / D_\text{os} = 1$ for the frequency $f = 7.2$~kHz, which coincides with the peak in Fig.~\ref{fil54_transverse_10mm}(b). Below this frequency, in the $0.1$~kHz to $5$~kHz range, most of the field is shielded by currents in the filaments, that are coupled via coupling currents, as discussed below.

Coupling losses are due to currents flowing between the filaments, known as the coupling currents~\cite{morgan1970theoretical}. They are represented by the arrows in Fig.~\ref{54_fil_transverse_sol}. It is worth mentioning that they are on average flowing anti-parallel to the applied magnetic field, as predicted by analytical models~\cite{morgan1970theoretical,campbell1982general}. Their dynamics is that of an RL-circuit governed by a time constant $\tau_\text{c}$ and they contribute to a loss per cycle and per unit length $q_\text{cycle}$ (J/m). Simplified models propose~\cite{campbell1982general}:
\begin{align}\label{eq_timeConstantCoupling}
\tau_\text{c} = \frac{\mu_0}{2\rho_{\text{eff}}} \paren{\frac{p}{2\pi}}^2, \quad q_\text{cycle} = \pi R_\text{w}^2\, \frac{b_\text{max}^2}{2\mu_0}\,\frac{\pi \omega \tau_\text{c}}{(\omega^2 \tau_\text{c}^2 + 1)}
\end{align}
with $\rho_{\text{eff}}$ the effective resistivity of the matrix, accounting for the presence of the filaments~\cite{wilson1983superconducting}. In the present case in which we assume no insulation between the filaments and the matrix, assuming the filaments have negligible resistivity, we can estimate $\rho_{\text{eff}}$ as follows~\cite{wilson2008nbti}:
\begin{align}
\rho_\text{eff} = \rho_\text{Cu}\, \frac{1-\lambda}{1+\lambda},
\end{align}
with $\lambda$ the filling factor of the filaments in the wire. Here, $\lambda = 0.44$ so that $\tau_\text{c} = 23$~ms. The associated frequency is $f_\text{c} = (2\pi \tau_\text{c})^{-1} = 7$~Hz, which roughly corresponds to the position of the peak value of the coupling loss per cycle in Fig.~\ref{fil54_transverse_10mm}(b). Below the peak frequency, the filaments are mostly decoupled, as the magnetic field does not change fast enough for large coupling currents to appear. Above the peak frequency, they get more and more coupled, as illustrated in Fig.~\ref{54_fil_transverse_sol_full_fil}.

\begin{figure}[h!]
        \centering
\begin{subfigure}[b]{0.99\linewidth}  
\centering
\tikzsetnextfilename{fil54_transverse_pitch_total}
\begin{tikzpicture}[trim axis left, trim axis right][font=\small]
\pgfplotsset{set layers}
 	\begin{loglogaxis}[
	tick label style={/pgf/number format/fixed},
    width=\linewidth,
    height=5cm,
    grid = both,
    grid style = dotted,
    ymin=1e-6, 
    ymax=1e2,
    xmin=0.01, 
    xmax=100000,
    ytick={1e-6, 1e-5, 1e-4, 1e-3, 1e-2, 1e-1, 1e0, 1e1, 1e2},
    ylabel={Average AC loss (W/m)},
    ylabel style={yshift=-2.5em},
    ylabel style={xshift=0em},
    xlabel style={yshift=0.4em},
    xticklabel style={yshift=-0.4em},
    yticklabel style={xshift=0em},
    yticklabel pos=right,
    legend columns=2,
    legend style={at={(1, 0.0)}, cells={anchor=west}, style={column sep=0cm}, inner sep=-0.5pt, anchor=south east, draw=none,fill opacity=0.5, text opacity = 1}
    ]
    \addplot[mag_6, thick] 
    table[x=f,y=total]{data/I_54fil_transverse_3mm_corr_scaled.txt};
        \addplot[mag_5, thick] 
    table[x=f,y=total]{data/I_54fil_transverse_10mm_corr_scaled.txt};
        \addplot[mag_4, thick] 
    table[x=f,y=total]{data/I_54fil_transverse_30mm_corr_scaled.txt};
        \addplot[mag_3, thick] 
    table[x=f,y=total]{data/I_54fil_transverse_100mm_corr_scaled.txt};
       	\addplot[black, densely dotted, thick] 
    table[x=f,y=total]{data/I_54fil_transverse_copper_scaled.txt};
        \addplot[mag_6, densely dashed] 
    table[x=f,y=coupling]{data/I_54fil_transverse_3mm_corr_scaled.txt};
        \addplot[mag_5, densely dashed] 
    table[x=f,y=coupling]{data/I_54fil_transverse_10mm_corr_scaled.txt};
        \addplot[mag_4, densely dashed] 
    table[x=f,y=coupling]{data/I_54fil_transverse_30mm_corr_scaled.txt};
        \addplot[mag_3, densely dashed] 
    table[x=f,y=coupling]{data/I_54fil_transverse_100mm_corr_scaled.txt};
    \legend{$p=3$ mm, $p=10$ mm, $p=30$ mm, $p=100$ mm, full copper}
    \node[] at (axis cs: 0.02, 20) {(a)};
    \end{loglogaxis}
\end{tikzpicture}%
\end{subfigure}
        \hfill
\begin{subfigure}[b]{0.99\linewidth}  
\centering
\tikzsetnextfilename{fil54_transverse_pitch_percycle}
\begin{tikzpicture}[trim axis left, trim axis right][font=\small]
\pgfplotsset{set layers}
 	\begin{loglogaxis}[
	tick label style={/pgf/number format/fixed},
    width=\linewidth,
    height=5cm,
    grid = both,
    grid style = dotted,
    ymin=1e-4, 
    xmin=0.01, 
    xmax=100000,
	xlabel={Frequency (Hz)},
    ylabel={AC loss per cycle (J/m)},
    ylabel style={yshift=-2.5em},
    xlabel style={yshift=0em},
    xticklabel style={yshift=-0.4em},
    yticklabel style={xshift=0em},
    yticklabel pos=right,
    legend columns=2,
    legend style={at={(0.5, 0.05)}, cells={anchor=west}, anchor=south, draw=none,fill opacity=0, text opacity = 1}
    ]
    \addplot[mag_6, thick] 
    table[x=f,y=total_f]{data/I_54fil_transverse_3mm_corr_scaled.txt};
        \addplot[mag_5, thick] 
    table[x=f,y=total_f]{data/I_54fil_transverse_10mm_corr_scaled.txt};
        \addplot[mag_4, thick] 
    table[x=f,y=total_f]{data/I_54fil_transverse_30mm_corr_scaled.txt};
        \addplot[mag_3, thick] 
    table[x=f,y=total_f]{data/I_54fil_transverse_100mm_corr_scaled.txt};
        \addplot[mag_6, densely dashed] 
    table[x=f,y=coupling_f]{data/I_54fil_transverse_3mm_corr_scaled.txt};
        \addplot[mag_5, densely dashed] 
    table[x=f,y=coupling_f]{data/I_54fil_transverse_10mm_corr_scaled.txt};
        \addplot[mag_4, densely dashed] 
    table[x=f,y=coupling_f]{data/I_54fil_transverse_30mm_corr_scaled.txt};
        \addplot[mag_3, densely dashed] 
    table[x=f,y=coupling_f]{data/I_54fil_transverse_100mm_corr_scaled.txt};
           	\addplot[black, densely dotted, thick] 
    table[x=f,y=total_f]{data/I_54fil_transverse_copper_scaled.txt};
    \node[] at (axis cs: 0.02, 0.015) {(b)};
    \end{loglogaxis}
\end{tikzpicture}%
\end{subfigure}
\vspace{-0.2cm}
\caption{AC loss as a function of the frequency of an applied transverse magnetic field for linear materials and different pitch lengths. Solid curves represent the total loss. Dashed curves represent the coupling loss only. The main effect of the twist is to shift the coupling loss curves to higher frequencies for decreasing values of $p$. The total loss for a wire made of copper only (2D model) is given for comparison (dotted curve). The legend is valid for both subfigures.}
        \label{fil54_transverse_pitch}
\end{figure}
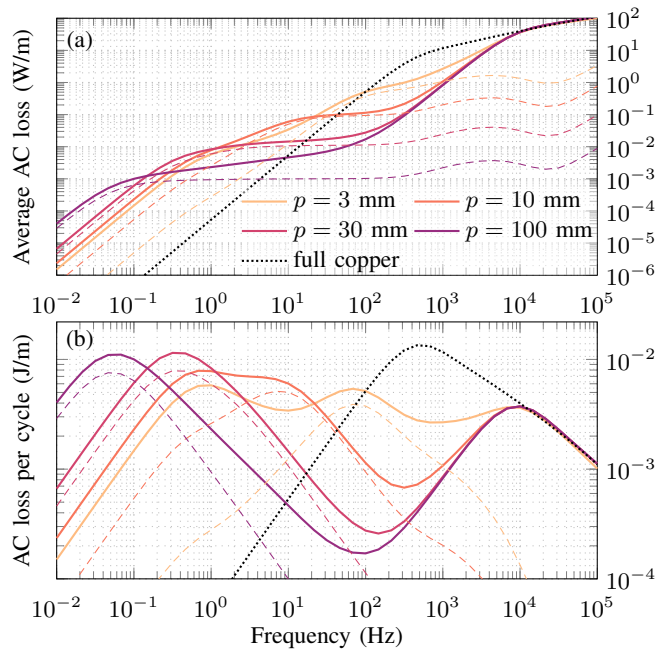

As an illustration of the effect of $p$ on the coupling losses, we give in Fig.~\ref{fil54_transverse_pitch} the total and coupling losses for different values of the pitch length. We can verify the agreement with the analytical prediction Eq.~\eqref{eq_timeConstantCoupling}: the peak position of the coupling losses scales quadratically with $p$, affecting the total loss significantly. The curve for a pure copper cylindrical conductor of the same radius $R_\text{w}$, with $\rho_\text{SC} = \rho_\text{Cu} = 1.81\times 10^{-10}$ $\Omega$m, is given for comparison.

Note that for the pure copper case, we have $\delta_{\text{Cu}}/R_\text{w} = 1$ for $f=183$~Hz, which roughly corresponds to the position of the peak of AC loss per cycle.

Figure~\ref{fil54_transverse_pitch} clearly shows the beneficial effect of twisting the filaments at low frequencies. Wires with smaller pitch length indeed have shorter time constants and are less subject to coupling current losses at low frequencies. It must be mentionned that the twist however does not reduce loss for all frequencies, which is in agreement with experimental measurements, e.g., in~\cite{kwasnitza1977scaling}.

The simple linear model discussed here allows for a qualitative description of coupling current losses that are representative of real superconducting wires for low applied magnetic field only. With superconducting filaments, saturation effects change the coupling current dynamics for higher field amplitudes~\cite{wilson1983superconducting}. Such effects cannot be reproduced with a linear model. 

As already said, the analysis of this linear model must therefore be carried out with caution. Total loss evaluations for superconducting wires cannot be extracted from this model as hysteresis losses of superconducting filaments are not part of the linear model. The inclusion of nonlinear material properties is necessary for such analysis. In the next section, we present the challenges that such an inclusion brings to the helicoidal transformation approach with general BC.


\subsection{Comments for nonlinear materials}\label{sec_commentNonlinear}

In the presence of nonlinear materials, such as superconducting filaments with a power law resistivity described by Eq.~\eqref{eqn_contitutiveje}, mode decoupling is no longer possible with general BC. As derived in the Appendix, the eddy current term of the formulation expands as a double sum on $k,k'\in \mathbb{Z}$ of the terms given by Eq.~\eqref{eq_longExpression}. And each term in Eq.~\eqref{eq_longExpression} involves the tensor $\rhotilde$. For a superconducting filament, this tensor depends on the full local current density, which couples the modes with different values of $|k|$. A large number of modes in Eq.~\eqref{eq_h_helicoidal_fullExpansion} is therefore likely to be excited by a transverse magnetic field.



As the integral along the $\xi_3$-direction can no longer be computed \textit{a priori}, the resulting problem is no longer two-dimensional, which makes it qualitatively different from the \tdpk\ model with linear materials. 

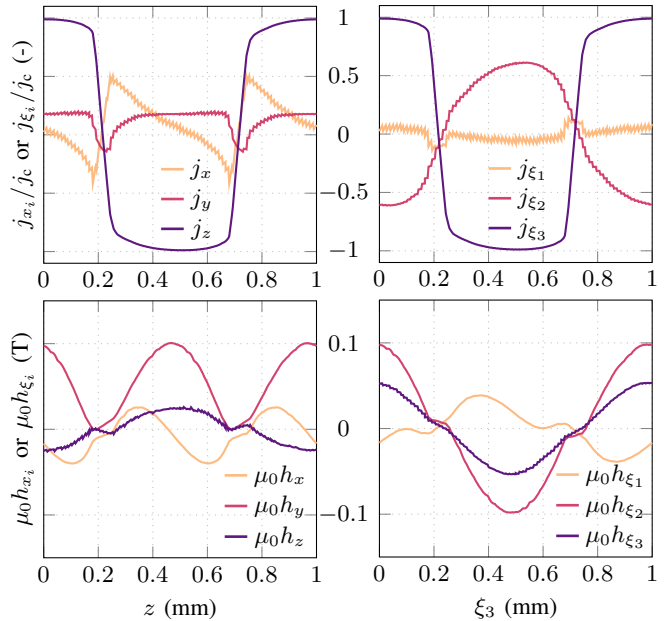
\begin{figure}[h!]
        \centering
        \begin{subfigure}[b]{0.49\linewidth}
        \centering
             \tikzsetnextfilename{filament_helicoidal_3D_transverse_super_x}
	\begin{tikzpicture}[trim axis left, trim axis right][font=\small]
\begin{groupplot}[group style={group size=1 by 2,
       horizontal sep=0pt,
       vertical sep=0.5cm},
     ] 
 	\nextgroupplot[
    width=1.2\linewidth,
    height=5cm,
    grid = both,
    grid style = dotted,
    xmin=0, 
    xmax=1,
    ymin=-1.1, 
    ymax=1.1,
    yticklabel=\empty,
	xtick scale label code/.code={},
	ytick scale label code/.code={},
    ylabel={$j_{x_i}/j_\text{c}$ or $j_{\xi_i}/j_\text{c}$ (-)},
    ylabel style={yshift=-3em},
    xlabel style={yshift=0.2em},
    legend style={at={(0.67, 0.25)}, cells={anchor=east}, anchor= east, draw=none, fill opacity=0,                 
    legend image code/.code={\draw[##1,line width=1pt] plot coordinates {(0cm,0cm) (0.3cm,0cm)};}, text opacity = 1}
    ]
    \addplot[mag_6, thick] 
    table[x=z,y=jx]{data/filament_helicoidal_3D_transverse_super.txt};
    \addplot[mag_4, thick] 
    table[x=z,y=jy]{data/filament_helicoidal_3D_transverse_super.txt};
        \addplot[mag_2, thick] 
    table[x=z,y=jz]{data/filament_helicoidal_3D_transverse_super.txt};
    \legend{$j_x$, $j_y$, $j_z$}
\nextgroupplot[
    width=1.2\textwidth,
    height=5cm,
    grid = both,
    grid style = dotted,
    xmin=0, 
    xmax=1,
    ymin=-0.15, 
    ymax=0.15,
    yticklabel=\empty,
	xtick scale label code/.code={},
    yticklabel style={anchor=center},
	ytick scale label code/.code={},
	xlabel={$z$ (mm)},
    ylabel={$\mu_0 h_{x_i}$ or $\mu_0 h_{\xi_i}$ (T)},
    ylabel style={yshift=-3em},
    xlabel style={yshift=0.3em},
    legend style={at={(0.82, 0.0)}, anchor=south, draw=none, fill opacity=0,                 
    legend image code/.code={\draw[##1,line width=1pt] plot coordinates {(0cm,0cm) (0.3cm,0cm)};}, text opacity = 1}
    ]
    \addplot[mag_6, thick] 
    table[x=z,y=bx]{data/filament_helicoidal_3D_transverse_super.txt};
    \addplot[mag_4, thick] 
    table[x=z,y=by]{data/filament_helicoidal_3D_transverse_super.txt};
        \addplot[mag_2, thick] 
    table[x=z,y=bz]{data/filament_helicoidal_3D_transverse_super.txt};
    \legend{$\mu_0 h_x$, $\mu_0 h_y$, $\mu_0 h_z$}
    \end{groupplot}
	\end{tikzpicture}%
        \label{filament_helicoidal_3D_transverse_super_x}
	        \end{subfigure}
        \begin{subfigure}[b]{0.49\linewidth}  
\centering
             \tikzsetnextfilename{filament_helicoidal_3D_transverse_super_xi}
	\begin{tikzpicture}[trim axis left, trim axis right][font=\small]
\begin{groupplot}[group style={group size=1 by 2,
       horizontal sep=0pt,
       vertical sep=0.5cm},
     ] 
  	\nextgroupplot[
    width=1.2\linewidth,
    height=5cm,
    grid = both,
    grid style = dotted,
    xmin=0, 
    xmax=1,
    ymin=-1.1, 
    ymax=1.1,
	xtick scale label code/.code={},
	ytick scale label code/.code={},
    ylabel style={yshift=-0.5em},
    xlabel style={yshift=0.2em},
    yticklabel style={anchor=center},
    yticklabel style={xshift=-1.5em},
    legend style={at={(0.67, 0.25)}, cells={anchor=east}, anchor= east, draw=none, fill opacity=0,
    legend image code/.code={\draw[##1,line width=1pt] plot coordinates {(0cm,0cm) (0.3cm,0cm)};}, text opacity = 1}
    ]
    \addplot[mag_6, thick] 
    table[x=z,y=jxi1]{data/filament_helicoidal_3D_transverse_super.txt};
    \addplot[mag_4, thick] 
    table[x=z,y=jxi2]{data/filament_helicoidal_3D_transverse_super.txt};
        \addplot[mag_2, thick] 
    table[x=z,y=jxi3]{data/filament_helicoidal_3D_transverse_super.txt};
    \legend{$j_{\xi_1}$, $j_{\xi_2}$, $j_{\xi_3}$}
\nextgroupplot[
    width=1.2\textwidth,
    height=5cm,
    grid = both,
    grid style = dotted,
    xmin=0, 
    xmax=1,
    ymin=-0.15, 
    ymax=0.15,
	xtick scale label code/.code={},
    yticklabel style={anchor=center},
	ytick scale label code/.code={},
	xlabel={$\xi_3$ (mm)},
    ylabel style={yshift=-0.5em},
    xlabel style={yshift=0.3em},
    yticklabel style={anchor=center},
    yticklabel style={xshift=-1.5em},
    legend style={at={(0.82, 0.0)}, anchor=south, draw=none, fill opacity=0,                 
    legend image code/.code={\draw[##1,line width=1pt] plot coordinates {(0cm,0cm) (0.3cm,0cm)};},
	text opacity = 1}
    ]
    \addplot[mag_6, thick] 
    table[x=z,y=bxi1]{data/filament_helicoidal_3D_transverse_super.txt};
    \addplot[mag_4, thick] 
    table[x=z,y=bxi2]{data/filament_helicoidal_3D_transverse_super.txt};
        \addplot[mag_2, thick] 
    table[x=z,y=bxi3]{data/filament_helicoidal_3D_transverse_super.txt};
    \legend{$\mu_0 h_{\xi_1}$, $\mu_0 h_{\xi_2}$, $\mu_0 h_{\xi_3}$}
    \end{groupplot}
	\end{tikzpicture}%
        \label{filament_helicoidal_3D_transverse_super_xi}
            \end{subfigure}
        \caption{Current density (up) and magnetic field (down) along the helicoidal fiber of pitch length $p$, passing through point $\vec x = \big(r, 0, 0\big)$, with $r=R_\ell + 0.8 R_\text{f}$ from $z=0$ to $p$, for a transverse applied magnetic field along $\ey$ and Nb-Ti filaments. (Left) Three components of the vectors in the $\vec x$-space. (Right) Three components of the vectors in the $\vec \xi$-space. Solution of the 3D model on a fine mesh with prismatic element in the filaments.}
        \label{filament_helicoidal_3D_transverse_super}
\end{figure}

To assess the importance of this mode coupling, we can use the 3D model. We show in Fig.~\ref{filament_helicoidal_3D_transverse_super} the evolution of the magnetic field and the current density along one helicoidal fiber, obtained with the 3D model with the same material parameters as in Section~\ref{sec_filaments_verif3Dmodel}, but subject to a transverse magnetic field. As can be seen on the bottom-right plot, the magnetic field in helicoidal coordinates cannot be described only with the two modes $f_{-1}(\xi_3)$ and $f_{+1}(\xi_3)$ as in the linear case. Higher modes are excited.

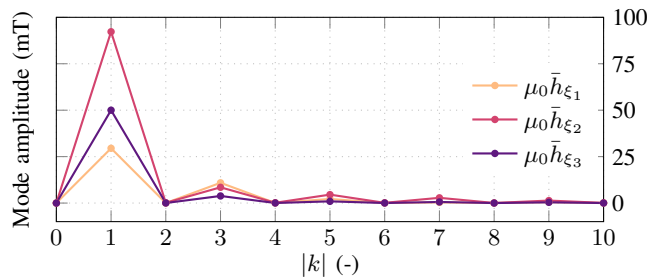
\begin{figure}[h!]
\centering
\tikzsetnextfilename{filament_helicoidal_3D_transverse_super_fft}
\begin{tikzpicture}[trim axis left, trim axis right][font=\small]
  \begin{axis}[
  	tick scale binop=\times,
    width=1\linewidth,
    height=4.3cm,
    grid = both,
    grid style = dotted,
    xmin=0,  
    xmax=10,
    ymax=0.1,
    xlabel={$|k|$ (-)},
    xtick=data,
    ytick={0, 0.025, 0.05, 0.075, 0.1},
    yticklabels={0, 25, 50, 75, 100},
    yticklabel pos=right,
    ylabel={Mode amplitude (mT)},
    ylabel style={yshift=-2.5em},
    xlabel style={yshift=0.7em},
    legend style={at={(0.99, 0.5)}, cells={anchor=east}, anchor=east, draw=none}
    ]
    \addplot[mag_6, thick, mark=*, mark options={mag_6, scale=0.5, style={solid}}] 
    table[x=k,y=b1k]{data/filament_helicoidal_3D_transverse_super_fft.txt};
    \addplot[mag_4, thick, mark=*, mark options={mag_4, scale=0.5, style={solid}}]  
    table[x=k,y=b2k]{data/filament_helicoidal_3D_transverse_super_fft.txt};
        \addplot[mag_2, thick, mark=*, mark options={mag_2, scale=0.5, style={solid}}] 
    table[x=k,y=b3k]{data/filament_helicoidal_3D_transverse_super_fft.txt};
    \legend{$\mu_0\bar h_{\xi_1}$, $\mu_0\bar h_{\xi_2}$, $\mu_0\bar h_{\xi_3}$}
    \end{axis}
\end{tikzpicture}%
\caption[Fourier mode amplitudes for the 3D model with a transverse field.]{Amplitudes of the mode contributions for the evolution of the three components of $\h$ in the $\vec \xi$-space, along the same helicoidal fiber as in Fig.~\ref{filament_helicoidal_3D_transverse_super}. Values are obtained via a fast Fourier transform. Modes for even numbers of $k$ are not excited by a transverse magnetic field.}
\label{filament_helicoidal_3D_transverse_super_fft}
\end{figure}

The amplitude of the different modes can be quantified by a discrete Fourier transform of the magnetic field evolution along this helicoidal fiber. This is illustrated in Fig.~\ref{filament_helicoidal_3D_transverse_super_fft}. Small, but non-negligible contributions are brought by modes $|k|>1$.

Whether a description with a limited number of modes would lead to satisfying evaluations of losses or not is not an obvious question; knowing a priori how many modes should be considered on a new geometry remains an open question. Further investigations in that direction are necessary.

\section{Conclusion}\label{sec_conclusion}

In this work, we applied a change of coordinates on the $h$\nobreakdash-$\phi$\nobreakdash-formulation for modelling multifilamentary wires presenting a helicoidal symmetry. This led to a reduction of the geometrical dimension from 3D to 2D, hence allowing for a substantial gain in terms of computational effort. We separated the study in two steps, depending on the helicoidal symmetry of the boundary conditions (BC). In both cases, we described in details the spatial discretization of finite element fields in helicoidal coordinates. In particular, we emphasized the necessity of using three independent components for the unknown fields. We then successfully verified our implementation against standard 3D models. 

In the case with no external field (e.g. transport current situation only) or with an axial magnetic field, BC are helicoidally symmetric and the method can be directly applied to nonlinear materials. The approach is exact in the sense that no approximation is introduced in the continuous setting. The proposed method can be directly applied on single-layer CORC\textsuperscript{\textregistered} cables~\cite{zhu2020advanced} or twisted stacked-tape conductors~\cite{takayasu2010cabling}.


In the case of a transverse magnetic field, BC are no longer helicoidally symmetric, but the approach was generalized and applied to linear materials. We presented a study of coupling current induced losses in the harmonic regime, and we finally commented on a possible extension to nonlinear materials. For nonlinear materials with general BC, further investigations are necessary.

\clearpage
\FloatBarrier
\section*{Appendix}

The full \hpf expressed in helicoidal coordinates reads: from an initial solution at $t=0$, find $\h \in\hsp(\O)$ such that, for $t>0$ and $\forall \h' \in \hspz(\O)$,
\begin{align}\label{eq_full_hpf_helicoidal_3D}
\volInt{\dt(\mutilde \, \h)}{\h'}{\Othreed} + \volInt{\rhotilde \ \curl \h}{\curl \h'}{\Octhreed} \qquad\notag\\
= \sum_{i\in C_V} \bar V_i \mathcal{I}_i(\h'),
\end{align}
with $\mutilde$ and $\rhotilde$ two tensors defined in Eqn.~\eqref{eq_matTensors} and \eqref{eq_matTensors_2}. This formulation is written in the 3D domain $\Othreed$, not yet reduced to a 2D problem. In the case of non-helicoidally symmetric boundary conditions, the solution $\h$ is not $\xi_3$-invariant, and the dimension reduction is not immediate.

Below, we expand the two integral terms of this formulation when the magnetic field $\h(\xi_1,\xi_2,\xi_3)$ is decomposed with Eq.~\eqref{eq_h_helicoidal_fullExpansion} and the modes $f_k(\xi_3)$ defined in Eq.~\eqref{eqn_modeDefinition}. We remind that these modes are orthonormal in the sense of the inner product defined in Eq.~\eqref{eq_innerProduct}.

\subsection*{Flux variation term (linear case)}

The first term of Eq.~\eqref{eq_full_hpf_helicoidal_3D} expands as the double sum
\begin{align}
\sum_{k=-\infty}^{\infty} \sum_{k'=-\infty}^{\infty}
&\volInt{\dt(\tilde{\vec \mu}\, \hpak{k}f_k)}{\hpak{k'}' f_{k'}}{\O_{\text{3D}}} \notag\\
&+ \volInt{\dt(\tilde{\vec \mu}\, \hpek{k}f_k)}{\hpak{k'}' f_{k'}}{\O_{\text{3D}}} \notag\\
&+\volInt{\dt(\tilde{\vec \mu}\, \hpak{k}f_k)}{\hpek{k'}' f_{k'}}{\O_{\text{3D}}}  \notag\\
&+ \volInt{\dt(\tilde{\vec \mu}\, \hpek{k}f_k)}{\hpek{k'}' f_{k'}}{\O_{\text{3D}}}.
\end{align}
Because the decomposition in Eq.~\eqref{eq_h_helicoidal_fullExpansion} separates the variables, we can integrate each individual term along the geometry invariant $\xi_3$-direction over one pitch length $p$. The orthogonality of the modes induces that terms with $k\neq k'$ vanish (provided that $\mu$ is not a function of the magnetic field). Dividing the integral by $p$ and using $\|f_k\| = 1$, we get,
\begin{align}
\sum_{k=-\infty}^{\infty} &\volInt{\dt(\tilde{\vec \mu}\, \hpak{k})}{\hpak{k}'}{\O} + \volInt{\dt(\tilde{\vec \mu}\, \hpek{k})}{\hpak{k}'}{\O} \notag \\
&+\volInt{\dt(\tilde{\vec \mu}\, \hpak{k})}{\hpek{k}'}{\O} + \volInt{\dt(\tilde{\vec \mu}\, \hpek{k})}{\hpek{k}'}{\O},
\end{align}
where integrals now only have to be performed on a 2D domain. Equations for different values of $|k|$ are uncoupled. In $\Occ$, the degrees of freedom for $\hpek{k}$ and $\hpak{-k}$ are linked with each other using Eq.~\eqref{eq_conditiononcoefficients}. 

\subsection*{Eddy current term (linear case)}

The second term of Eq.~\eqref{eq_full_hpf_helicoidal_3D} expands as the double sum
\begin{align}
\sum_{k=-\infty}^{\infty} &\sum_{k'=-\infty}^{\infty}
\volInt{\tilde{\vec \rho}\, \curl(\hpak{k}f_k)}{\curl(\hpak{k'}' f_{k'})}{\O_{\text{c,3D}}} \notag\\
&\quad+ \volInt{\tilde{\vec \rho}\, \curl(\hpek{k}f_k)}{\curl(\hpak{k'}' f_{k'})}{\O_{\text{c,3D}}} \notag\\
&\quad+\volInt{\tilde{\vec \rho}\, \curl(\hpak{k}f_k)}{\curl(\hpek{k'}' f_{k')}}{\O_{\text{c,3D}}} \notag\\
&\quad+ \volInt{\tilde{\vec \rho}\, \curl(\hpek{k}f_k)}{\curl(\hpek{k'}' f_{k'})}{\O_{\text{c,3D}}}.
\end{align}
Using Eq.~\eqref{eq_curlDecomposition} for the curl, we get the following lengthy expression for each pair of values $(k,k')\in \mathbb{Z}\times\mathbb{Z}$,
\begin{align}\label{eq_longExpression}
&\volInt{\tilde{\vec \rho}\, f_k\ \curl \hpak{k}}{f_{k'}\ \curl \hpak{k'}'}{\O_{\text{c,3D}}}\notag\\
+& \volInt{\tilde{\vec \rho}\, \derd{f_k}{\xi_3} \ezh \times \hpak{k}}{f_{k'}\ \curl \hpak{k'}'}{\O_{\text{c,3D}}}\notag\\
+&\volInt{\tilde{\vec \rho}\, f_k\ \curl \hpak{k}}{\derd{f_{k'}}{\xi_3}\ezh \times \hpak{k'}'}{\O_{\text{c,3D}}}\notag\\
+& \volInt{\tilde{\vec \rho}\, \derd{f_k}{\xi_3} \ezh \times \hpak{k}}{\derd{f_{k'}}{\xi_3} \ezh \times \hpak{k'}'}{\O_{\text{c,3D}}}\notag\\
+& \volInt{\tilde{\vec \rho}\, f_k \curl \hpek{k}}{f_{k'}\ \curl \hpak{k'}'}{\O_{\text{c,3D}}}\notag\\
+& \volInt{\tilde{\vec \rho}\, f_k \curl \hpek{k}}{\derd{f_{k'}}{\xi_3} \ezh \times \hpak{k'}'}{\O_{\text{c,3D}}}\notag\\
+& \volInt{\tilde{\vec \rho}\, f_k\ \curl \hpak{k}}{f_{k'} \curl \hpek{k'}'}{\O_{\text{c,3D}}}\notag\\
+& \volInt{\tilde{\vec \rho}\, \derd{f_k}{\xi_3} \ezh \times \hpak{k}}{f_{k'} \curl \hpek{k'}'}{\O_{\text{c,3D}}} \notag\\
+& \volInt{\tilde{\vec \rho}\, f_k \curl \hpek{k}}{f_{k'} \curl \hpek{k'}'}{\O_{\text{c,3D}}}.
\end{align}
In the linear case in which $\rhotilde$ is not a function of the fields, we can integrate each term along the geometry invariant $\xi_3$-direction over one pitch length $p$, divide by $p$, use the mode property Eq.~\eqref{modesProperty}, and exploit the mode orthonormality.

For $k=0$, because $d_{\xi_3}f_0 = 0$, only terms for $k'=0$ survive, and they are decoupled from all other terms ($k\neq 0$). These terms are the same as the ones implemented in the case of helicoidally symmetric boundary conditions (HS-BC):
\begin{align}
&\volInt{\tilde{\vec \rho}\, \curl \hpak{0}}{\curl \hpak{0}'}{\Oc} +  \volInt{\tilde{\vec \rho}\, \curl \hpek{0}}{\curl \hpak{0}'}{\Oc}\notag \\
 +& \volInt{\tilde{\vec \rho}\, \curl \hpak{0}}{\curl \hpek{0}'}{\Oc} + \volInt{\tilde{\vec \rho}\, \curl \hpek{0}}{\curl \hpek{0}'}{\Oc}.
\end{align}

For $k\neq 0$, only one term of the sum on $k'$ survives for each term, either $k'=k$, or $k'=-k$. Indeed, Eq.~\eqref{modesProperty} induces the coupling of the modes $k$ and $-k$. For a given value of $k\neq 0$, in Eq.~\eqref{eq_longExpression}, the only terms that remain are
\begin{align}
& \volInt{\tilde{\vec \rho}\, \curl \hpak{k}}{\curl \hpak{k}'}{\Oc}\notag\\
+& \alpha k \volInt{\tilde{\vec \rho}\, \ezh \times \hpak{k}}{\curl \hpak{-k}'}{\Oc}\notag\\
+&\alpha (-k) \volInt{\tilde{\vec \rho}\, \curl \hpak{k}}{\ezh \times \hpak{-k}'}{\Oc}\notag\\
+& \alpha^2 k^2 \volInt{\tilde{\vec \rho}\, \ezh \times \hpak{k}}{\ezh \times \hpak{k}'}{\Oc}\notag\\
+& \volInt{\tilde{\vec \rho}\, \curl \hpek{k}}{\curl \hpak{k}'}{\Oc}\notag\\
 +& \alpha (-k) \volInt{\tilde{\vec \rho}\, \curl \hpek{k}}{\ezh \times \hpak{-k}'}{\Oc}\notag\\
+& \volInt{\tilde{\vec \rho}\, \curl \hpak{k}}{\curl \hpek{k}'}{\Oc}\notag\\
 +& \alpha k \volInt{\tilde{\vec \rho}\, \ezh \times \hpak{k}}{\curl \hpek{-k}'}{\Oc} \notag\\
+& \volInt{\tilde{\vec \rho}\, \curl \hpek{k}}{\curl \hpek{k}'}{\Oc}.
\end{align}
To these terms, another set needs to be added, with the opposite value of $k$, $k^\star = -k$. In total, this gives eighteen individual terms for the eddy current contribution, for each value of $|k|\neq 0$.

%
%

\bibliographystyle{ieeetr}
\bibliography{../paperReferences}

\end{document}